\def\year{2019}\relax
\documentclass[letterpaper]{article} 
\usepackage{aaai19}  
\usepackage{times}  
\usepackage{helvet}  
\usepackage{courier}  
\usepackage{url}  
\usepackage{mdwmath}
\usepackage{mdwtab}
\usepackage{mathrsfs}
\usepackage{microtype}
\usepackage{amsmath,amssymb,amsthm}
\usepackage{graphicx}  
\usepackage{xcolor}
\usepackage[ruled,vlined,linesnumbered]{algorithm2e}
\usepackage[capitalize]{cleveref}

\newcommand{\xhdr}[1]{\vspace{0.25mm}\noindent{{\bf #1.}}}
\newcommand{\hide}[1]{}

\begin{document}
%
\title{Modeling and Analysis of Tagging Networks in Stack Exchange Communities}
\author{
Xiang Fu\thanks{The first two authors contributed equally to this work.},\ and  Shangdi Yu\footnotemark[1],\ and Austin R.~Benson\\
Computer Science Department\\
Cornell University\\
\{xf74, sy543\}@cornell.edu,\ arb@cs.cornell.edu \\
}

\maketitle
\begin{abstract}
Large Question-and-Answer (Q\&A) platforms support diverse knowledge curation on
the Web.
While researchers have studied user behavior on the platforms in a variety of
contexts, there is relatively little insight into important by-products of user
behavior that also encode knowledge.
Here, we analyze and model the macroscopic structure of tags applied by users to
annotate and catalog questions, using a collection of 168 Stack Exchange
websites.
We find striking similarity in tagging structure across these Stack Exchange
communities, even though each community evolves independently (albeit under
similar guidelines).
Using our empirical findings, we develop a simple generative model that creates
random bipartite graphs of tags and questions.
Our model accounts for the tag frequency distribution but does not explicitly account for
co-tagging correlations.
Even under these constraints, we demonstrate empirically and theoretically that
our model can reproduce a number of statistical
properties of the co-tagging graph that links tags appearing in the same post.
\end{abstract}


\section{Introduction}

Question-and-Answer (Q\&A) platforms are now a standard context for social interaction
on the Web with platforms such as Quora and Stack Exchange supporting large user bases.
As a result, the social networks that these platforms support
have undergone a great deal of study, including, for example,
how people find interesting and popular questions on Quora~\cite{Wang2013},
prediction of ``best answer'' selection on Yahoo Answers~\cite{yahoo1},
market design for knowledge base construction with Google Answers~\cite{Chen-2010-Google}, and
badge collection on Stack Overflow~\cite{Anderson-2013-badges}.
These studies have largely focused on models and analysis of the user behavior.
However, the users also create other types of richly structured data. In
this paper, we model and analyze the structure revealed by tags on Stack
Exchange, which are used to annotate and catalog questions. Thus, our principal
object of study is the tags (and their relationships through co-tagging), rather
than the users; however, tags are still a by-product of user behavior since
users apply the tags.

\begin{figure}[tb]
\centering
\includegraphics[width=0.8\columnwidth]{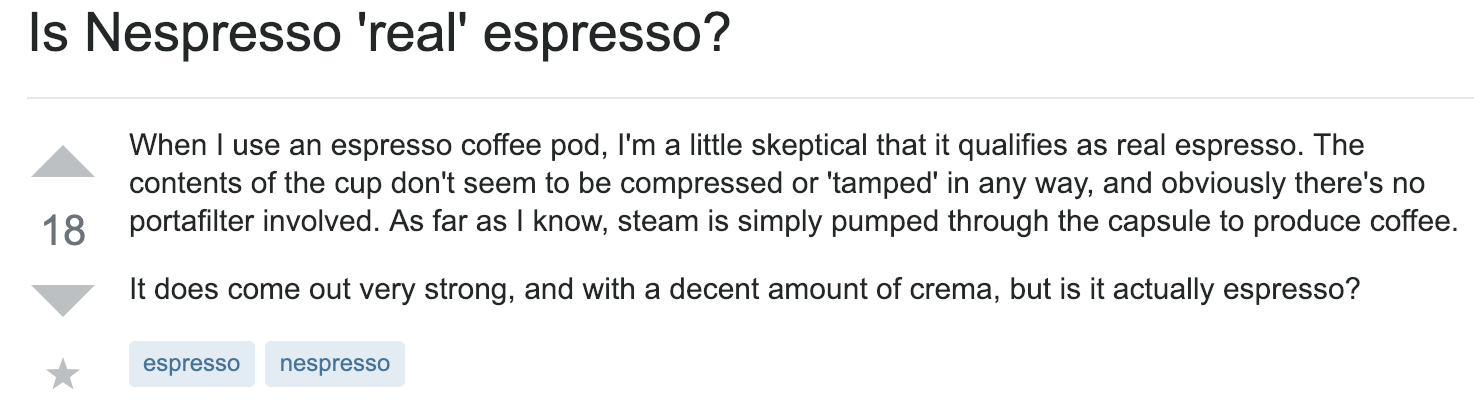}
\includegraphics[width=0.5\columnwidth]{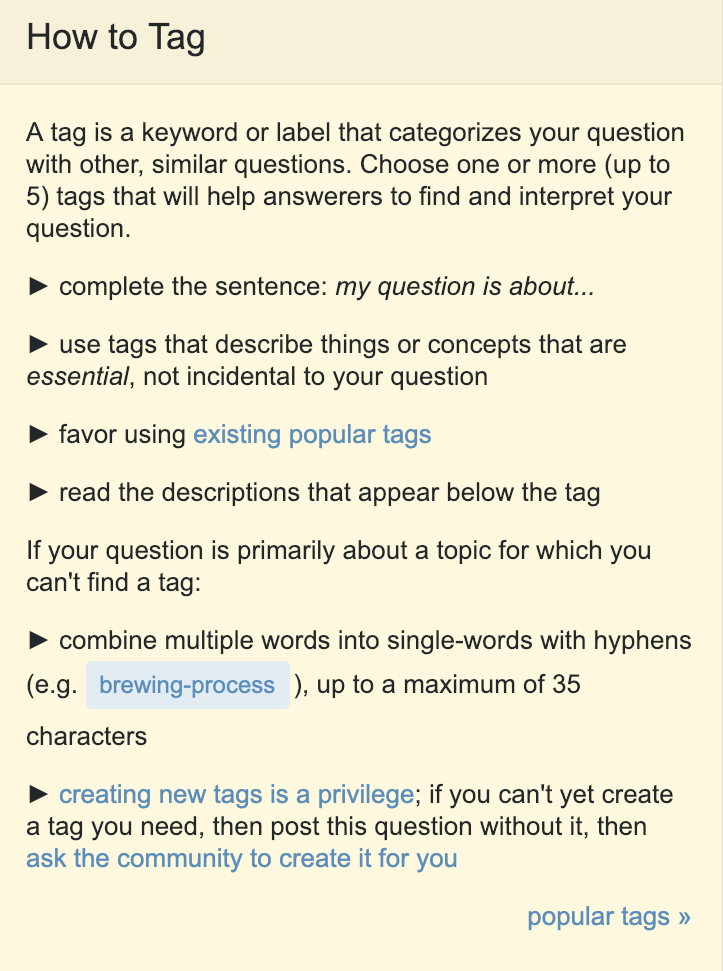}
\caption{Stack Exchange tagging. {\bf (Top)} A question on the \textsc{coffee} Stack Exchange
  community with two tags: \texttt{espresso} and \texttt{nespresso}
  (\url{https://coffee.stackexchange.com/q/1572}). We study the tag frequency distributions
  across a large collection of Stack Exchange communities, as well as networks
  constructed from tags applied to the same questions.
  {\bf (Bottom)} User interface of tagging guidelines on the \textsc{coffee} Stack Exchange
  (\url{https://coffee.stackexchange.com/questions/ask}). The last rule says
  that users cannot immediately create new tags without due process; thus, tagging
  is fundamentally different from hashtags on other social media platforms such as
  Twitter or Instagram.
}
\label{fig:coffee_sx}
\end{figure}

A Stack Exchange website is a Q\&A forum for a particular
community. The platform began with Stack Overflow, which is a community for computer
programming. Stack Overflow is the the largest and arguably most well-known
Stack Exchange community, but the Stack Exchange ecosystem supports a diverse
set of communities ranging from
pet ownership\footnote{https://pets.stackexchange.com/} to
coffee\footnote{https://coffee.stackexchange.com/} to
philosophy.\footnote{https://philosophy.stackexchange.com/}
For the most part, these Stack Exchange communities evolve independently under the
same Q\&A format (\cref{fig:coffee_sx}).
A linchpin of every Stack Exchange community is the tagging system.
When posting a question, users are encouraged to apply a small
number of tags (at least one and at most five) that provide
a reasonable abstraction of the question's topics.
In addition to describing the question's content, tags also serve
users in information retrieval of similar questions as well as questions they might be able to answer.
Tags on Stack Exchange are not taken lightly---users cannot immediately create new
tags and are encouraged to use existing and popular tags (\cref{fig:coffee_sx}, bottom);
moreover, there are also official tagging guidelines.\footnote{https://stackoverflow.com/help/tagging}
Thus, tags on Stack Exchange are fundamentally different from, e.g., hashtags on social
media platforms such as Twitter which are largely free from regulation.
The value placed on tags means that they can contain rich information about the community.
For example, tag frequencies can show popular topics and the change of tag
frequency over time can reveal the change of a community's interests over time.

Here, we provide the first large-scale study of the macroscopic structure of
tagging behavior by analyzing a collection of 168 Stack Exchange communities. We
frame our study through the lens of network analysis, focusing on two networks
constructed from the tagging behavior of users. The first is the bipartite
network of tags and questions, where there is an edge between a tag and all of
the questions to which the tag was applied. The second is the co-tagging
network, or the projection of the first network onto the tags; in this case, two
tags are connected by an edge if the two tags jointly annotate at least one
question. (We also consider a weighted version of the second network, where the
weight is the number of questions containing the two tags.)

Oftentimes, network analyses suffer from the fact that there is only ``one sample''
of a social system to study. For example, there is only one Facebook
friendship graph~\cite{Ugander-2011-anatomy} and one Twitter follower
network~\cite{Kwak-2010-Twitter} to analyze. While such studies provide valuable
insights into real-world social systems, it is also well-known that there can
be randomness in the evolution of social networks when
crafted in a controlled setting~\cite{Salganik-2006-MusicLab}. The Stack
Exchange communities provide a unique opportunity to study a collection of
similar networks of tags with highly similar dynamics that have evolved
largely independently and differ most in the community topic
(and implications of the community topic, such as the number of users).

We begin with an empirical analysis on the frequency distribution of the tags
across our collection of 168 Stack Exchange communities. We find that this distribution is
heavy-tailed and well-approximated by a lognormal distribution, and the two
parameters of this distribution are themselves well-approximated by a normal
distribution when estimated over the large collection of Stack Exchange communities.
From our findings, we devise a simple generative model for creating random bipartite
graphs with links connecting tags to questions. The model takes as input the
desired number of questions, number of tags, number of total tag occurrences, and
two parameters of a lognormal distribution, and produces as output a
bipartite graph linking tags to questions.

We futher explore the Stack Exchange data by analyzing the ``co-tagging network''
induced by the bipartite tag-question network. Specifically, we analyze the
graph where the nodes are tags and there is an edge connecting two tags if they
are ``co-tagged'' on at least one question (with possible weighting on edges
corresponding to the number of questions on which the two tags appear). Our
analysis focuses on three macroscopic properties of the data.  First, the
weighted number of co-tags of a given tag is well-approximated by a linear
function of the number of questions in which the tag appears. Second, the number
of unique co-tags of a given tag is well-approximated by a simple third-degree
polynomial of the number of the tag frequency. Qualitatively, as we increase the number of
questions that a tag has appeared in, the number of unique co-tags will also
increas; however, this growth tapers for popular tags, when it is difficult to
accumulate more unique co-tags. Third, we measure three versions of the
clustering coefficient for weighted and unweighted networks networks and find
various levels of clustering and find that the unweighted clustering coefficient
is only mildly correlated with the size of the Stack Exchange community (as
measured by the number of questions), but two versions of the weighted versions
both negatively correlate with size.

All three macroscopic properties are replicated by our model, which we validate
with both empirical and theoretical analysis across the collection of 168 Stack
Exchange networks. Importantly, the model does not bake in any notions of
correlation or clustering in the co-tagging but can still replicate important
co-tagging network properties. Thus, we can conclude that these network
properties could actually be simply explained by our simple generative model that only
makes a strong assumption on the frequency distribution of the tags. These findings
contrast sharply with traditional social network analysis in measuring clustering.
Standard random graph models for social networks that do not bake in clustering structure
do not exhibit the same clustering levels as the real-world social system~\cite{Newman-2003-survey}.
However, in our case, the co-tagging network constructed from our bipartite tag-question
generative model matches the clustering levels in the empirical data.

\section{Related Work}

We summarize below how our research relates to several areas in social media, information
retrieval, and network science.

\subsubsection{Online Q\&A platforms}
Question-and-Answer (Q\&A) platforms have been a staple of online discussion
for several years, involving major web companies such as Yahoo!, Google, and Quora.
Research on these platforms has spanned a variety of topics, including
reputation mechanisms~\cite{rep1,rep2},
answer quality measurement~\cite{Wang2013,stackE,Anderson-2012-value},
network structure~\cite{yahoo1,Paranjape-2017-motifs};
social behavior~\cite{yang2011culture};
answer prediction~\cite{yahoo1,tian2013towards};
topic popularity~\cite{topic_pop}; and
expertise evaluation~\cite{exper1,stackE,Pal-2012-evolution}.
This research has largely focused on the questions, answers, and user behavior.
Our paper, in contrast, treats tags as the fundamental object of
study. Furthermore, most prior work has only examined at most a few Q\&A web
sites, whereas we study a large collection of Stack Exchange networks.

\subsubsection{Folksonomy}
The tag-question network that we study is related to the idea of \emph{folksonomy},
a term coined by Thomas Vander Wal to describe the practice of users
tagging information for personal retrieval in an open social
environment~\cite{vander}. Folksonomy has been a lens for analysis on
social media platforms such as \texttt{CiteULike}, \texttt{del.icio.us}, and
\texttt{BibSonomy}~\cite{Cattuto-2007-folksonomy,citeulike,Cattuto-2009-folksonomy}.
A major difference of these folksonomy studies and the present work is that
folksonomies are much less restricted in the annotations---users can add many
(possibly new) annotations freely---whereas the Stack Exchange system is
restricted (between one and five tags with systematic vetting of new tags).
And again, we analyze a large collection of Stack Exchange communities and not
just a few folksonomies.

\subsubsection{Bipartite network models and co-tagging networks}
Bipartite graph (network) models are employed across a broad range of scientific
disciplines, including
ecology~\cite{Bascompte-2006-coevolution},
biomedicine~\cite{Goh-2007-disease}, and
information science~\cite{Akoglu-2013-opinion}.
The model that we develop in this paper is
a \emph{generative (random) model} for a bipartite graph (network) between tags and questions.
Other generative models for bipartite (or multipartite) graphs include the
bipartite stochastic block model~\cite{Larremore-2014-biSBM},
evolutionary affiliation networks~\cite{Lattanzi-2009-affiliation}, and
generative models for folksonomy~\cite{Chojnacki-2010-folksonomy}.
In contrast to prior research, the goal with our model is to develop a simple
generative model that captures the empirical properties that we observe to
persist across Stack Exchange communities. Our model is designed to capture the
tag frequency distribution amongst questions, but we find that properties of the
co-tagging network---where tags are connected if they have appeared in a
question together---are still replicated with our model.
Properties and statistics of co-tagging networks, such as clustering
coefficients, characteristic path lengths, and number of co-tags have been used
to analyze online communities such as \texttt{del.icio.us} and \texttt{BibSonomy},
have been studied~\cite{Cattuto-2007-folksonomy,cotag1}. Co-tagging networks have also
been used for application on connecting users with similar interests~\cite{cotag2}.

\section{Data Description and Preliminary Analysis}
A Stack Exchange is a self-moderating online Q\&A forum, and
each Stack Exchange community centers on a different topic. Questions
are annotated with at least one and at most five tags that serve as essential
descriptors of the question (\cref{fig:coffee_sx}). Importantly, these platforms
also largely evolve independently, allowing us to perform a better statistical analysis
compared to analyzing a single Stack Exchange community.
We now describe our dataset collection and provide preliminary statistical analyses that will serve the
development of our generative model in the next section.

\begin{figure}
\centering
\includegraphics[width=0.9\columnwidth]{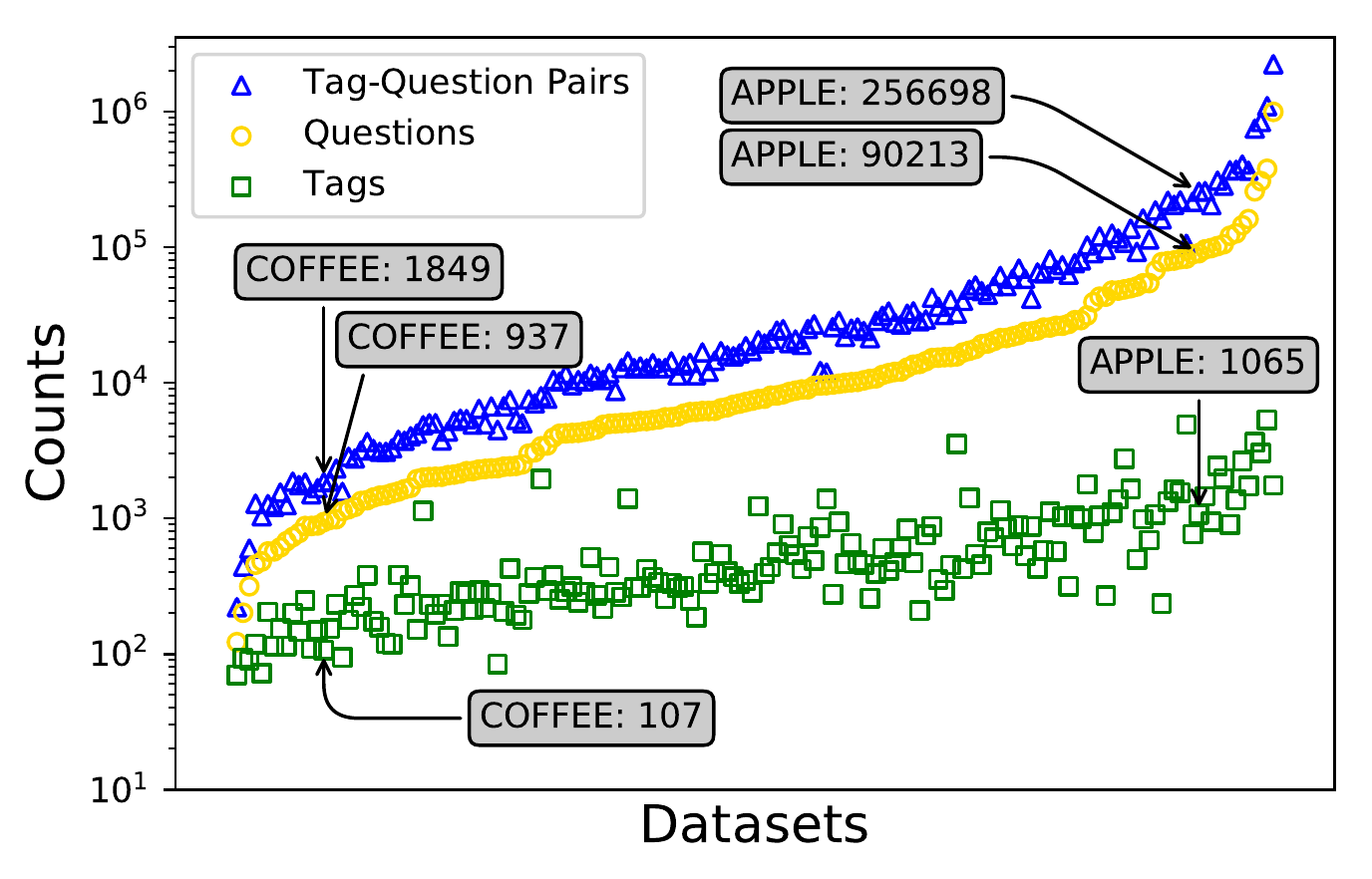}
\caption{Number of unique tags (green squares), number of questions (yellow circles), and tag-question pairs (blue triangles) of
the 168 Stack Exchange communities that we analyze in this paper. Datasets are sorted in ascending order by number of questions.
The \textsc{apple} and \textsc{coffee} communities are annotated as examples.
In this paper, we analyze and model the relationships between tags and questions.
}
\label{fig:data_sta}
\end{figure}

\subsection{Data Collection}
We collected data from \url{https://archive.org/details/stackexchange}, which hosts
the entire history of every Stack Exchange community, including the tags used
to annotate questions.
In total, we collected the sets of tags applied to each question from 168
Stack Exchange communities. In order to ensure that we could analyze
data by inspection, we omitted communities where the
predominant language was not English (thus, we do not consider the
\textsc{es}, \textsc{ja}, \textsc{pt}, \textsc{ru}, \textsc{rus}, and \textsc{ukrainian} communities).
However, we do include Stack Exchange communities such as \textsc{russian},
where people discuss the Russian language in English.
We also omitted so-called ``meta'' communities that discuss
a particular Stack Exchange community since these meta communities have a different set of goals
as well as a dependence on the community that they discuss.
Finally, we also omitted Stack Overflow, which is over an order of magnitude larger than any other
community, and has already been the subject of much research~\cite{Anderson-2012-value,rep1,Wang2013}.
We will release our processed dataset collection with the publication of this paper.

\Cref{fig:data_sta} presents an overview of the basic statistics of our collection of tags.
Among the 168 Stack Exchange communities that we study in this paper,
the number of unique tags ranges from 70 (\textsc{arabic}) to 5,318 (\textsc{superuser}),
and the number of questions ranges from 122 (again, \textsc{arabic}) to 994,983 (\textsc{math}).
Although the Stack Exchange communities vary in size and topic and also evolve largely
independently, we see in the next section (and later in the paper) that there are broad similarities
across the communities.

 \begin{figure}
\centering
\includegraphics[width=1.0\columnwidth]{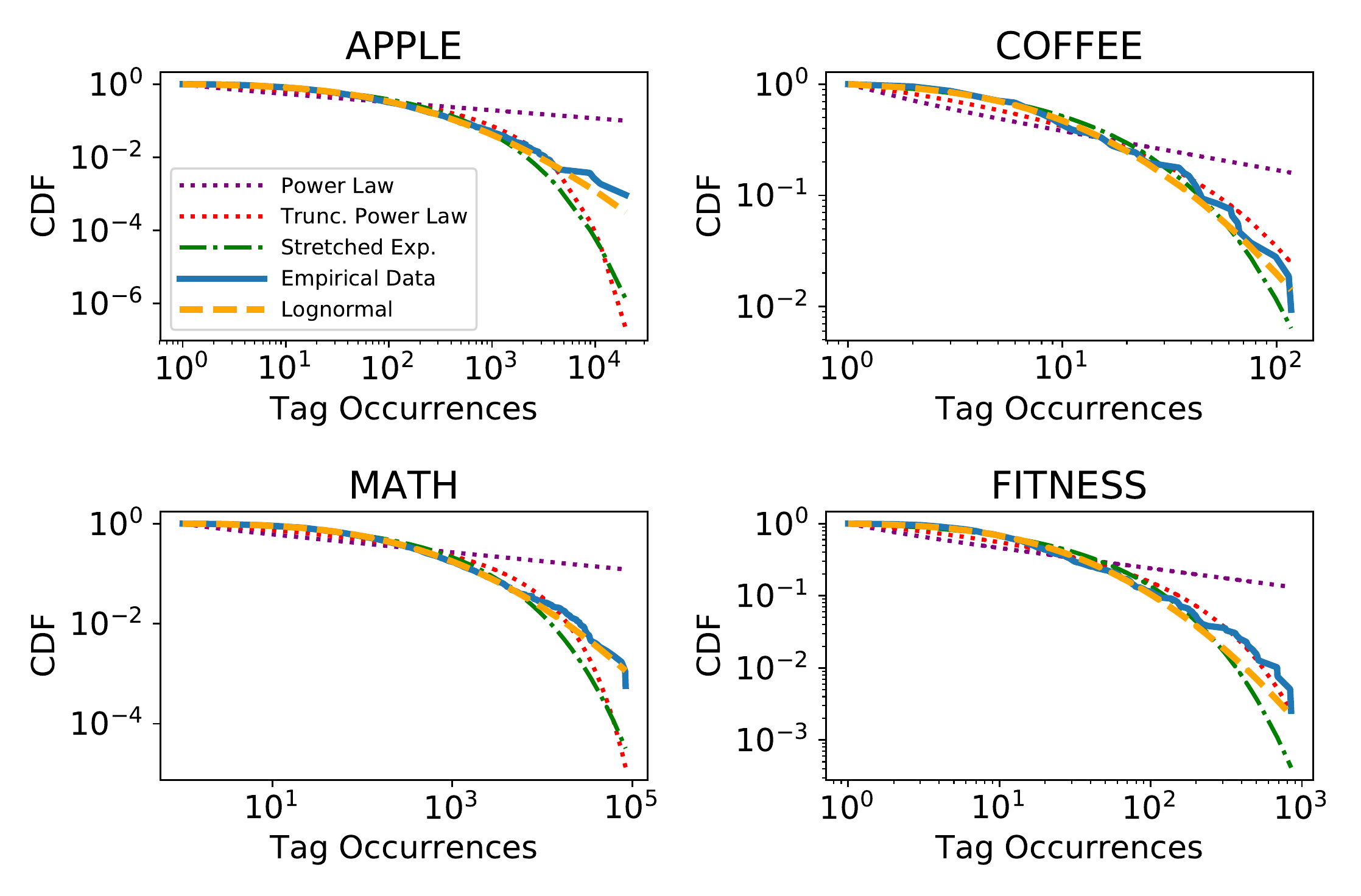}
\caption{Tag frequencies of four diverse Stack Exchange communities
    (\textsc{apple}---90,213 questions and 1,065 tags;
     \textsc{coffee}---937 questions and 107 tags;
     \textsc{math}---994,983 questions and 1,749 tags;
     \textsc{fitness}---7,626 questions and 393 tags).
     We find that tag frequencies are well-modeled by a lognormal distribution in Stack Exchange
     communities (\cref{fig:ks}).
     }
\label{fig:tag_freq}
\end{figure}

\subsection{Lognormal Distribution of Tag Frequencies}
In this section, we study the distribution of tag frequencies,
i.e., the number of times that a tag is applied to a question or, when normalized, the fraction of questions that contains a given tag.
One consistent trait is that tags used only a few times are much more common than tags used many times,
and the distribution of tags is heavy-tailed. Many
communities have tags appearing at much higher frequencies than most other tags;
as an extreme example, the \texttt{magic-the-gathering} tag appears in more than 3000 questions in the \textsc{boardgames} community,
while all other tags appear in fewer than 500 questions.

Such heavy-tailed distributions are common on the Web and other
domains~\cite{Mitzenmacher-2004-history,power-law}.
Here, we find that the tag frequencies are well-modeled by a lognormal distribution.
\Cref{fig:tag_freq} illustrates four representative cases and also provides a comparison against other commonly-used
heavy-tailed probability distributions such as a power law, truncated power law, and stretched exponential.
(\cref{fig:tag_freq} shows four such cases). We find that a lognormal tends to match both the head and tail
of the distribution, while other common heavy-tailed distributions can only capture either the head or tail
of the distribution (e.g., in \cref{fig:tag_freq}, the truncated power law captures the head of the tag frequency
distribution in \textsc{apple} but not the tail and the tail of the \textsc{coffee} distribution but not the head).
The lone outlier is the \textsc{patent} community, which does not seem to be well-approximated by any
commonly-used heavy-tailed distribution.

\begin{figure}
    \centering
    \includegraphics[width=0.495\columnwidth]{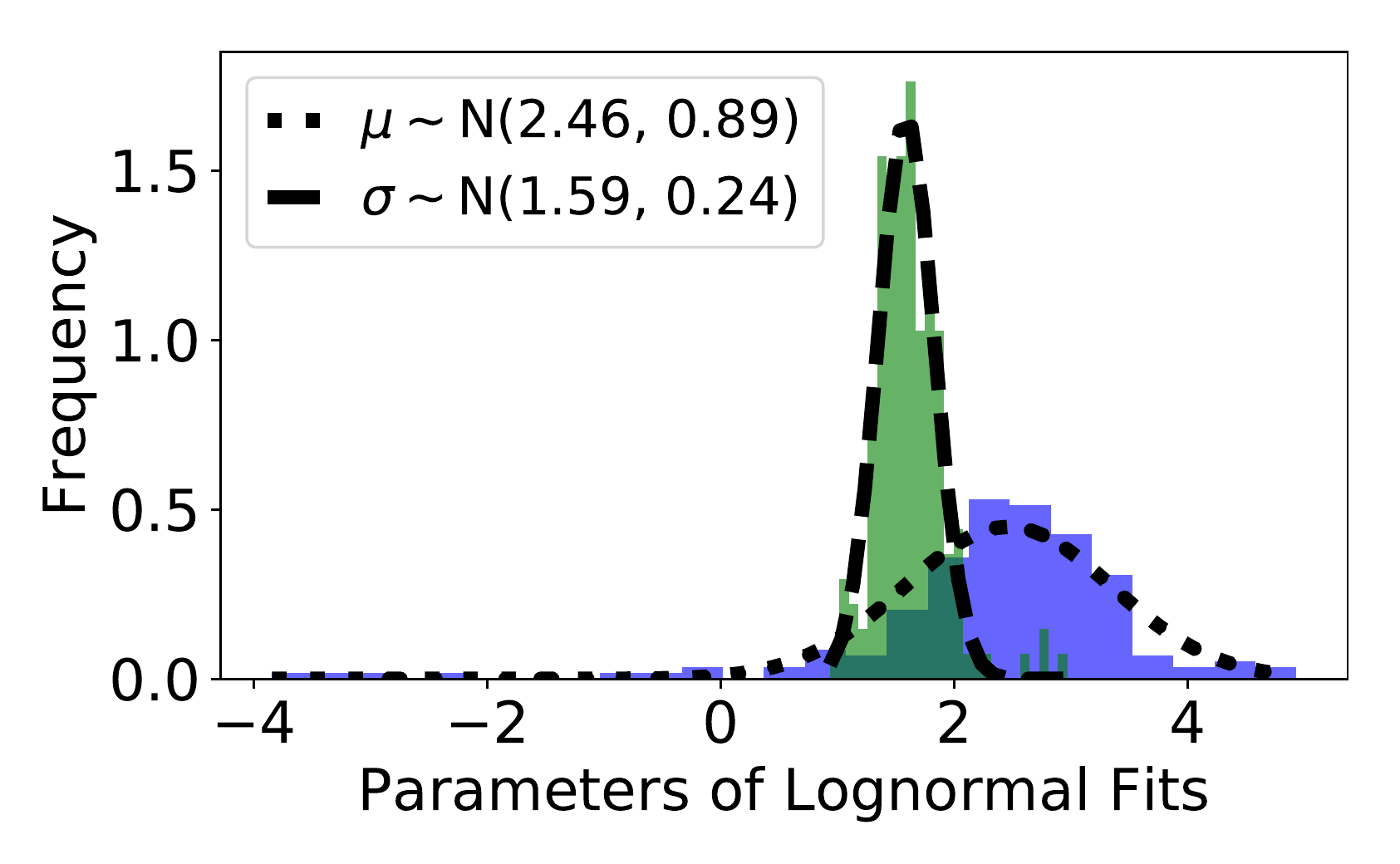}
    \includegraphics[width = 0.495\columnwidth]{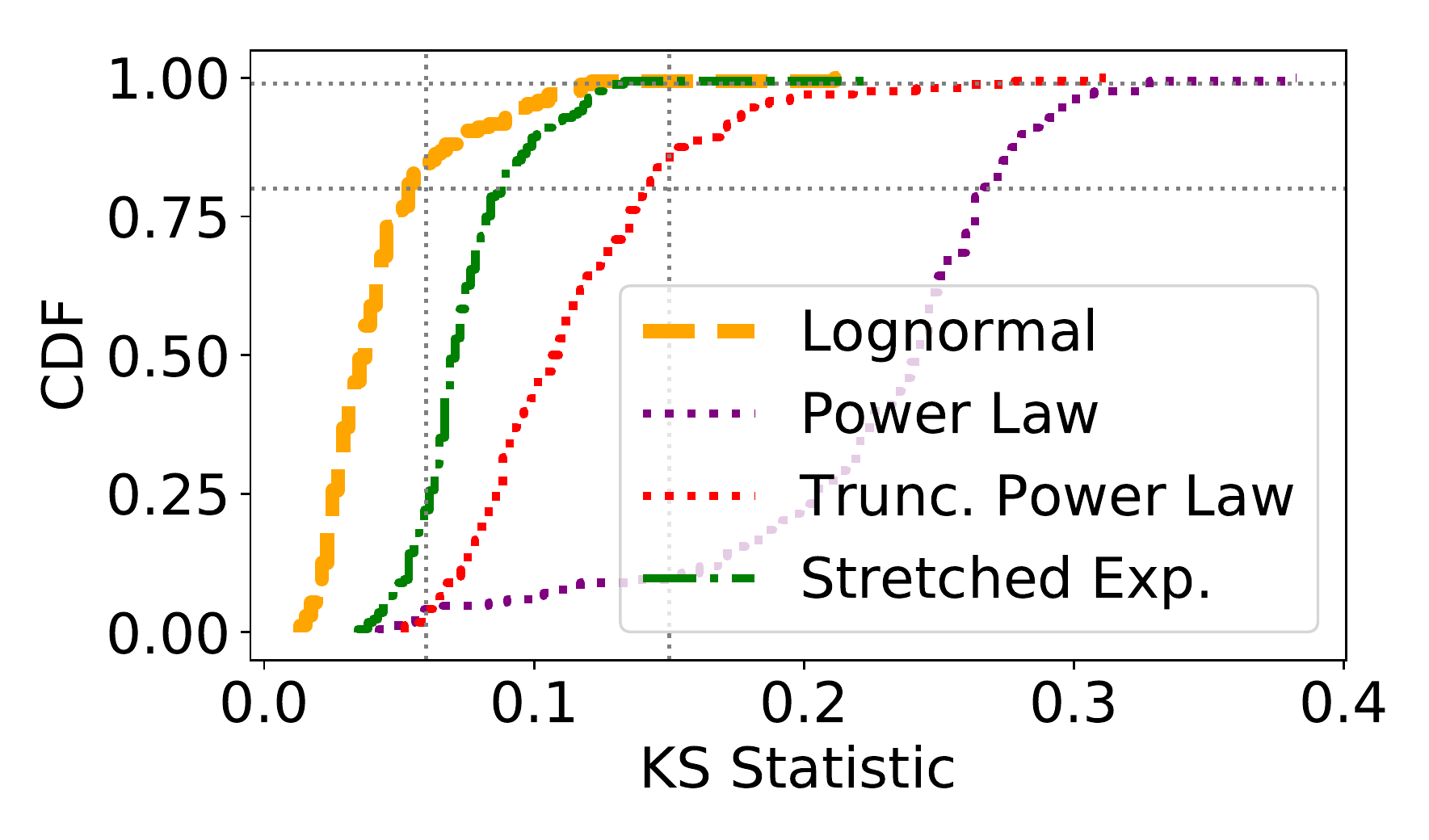}
     \includegraphics[width=0.495\columnwidth]{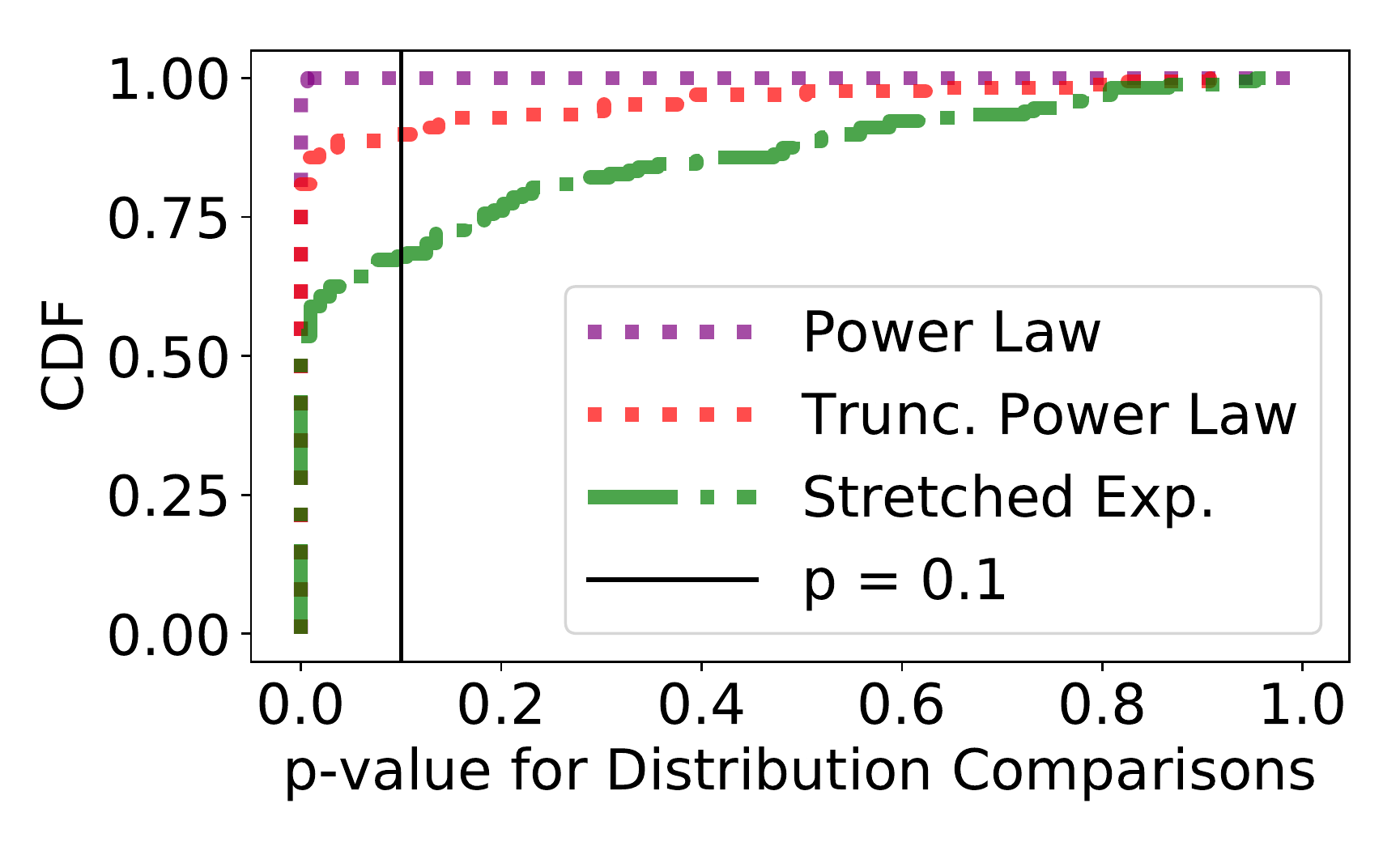}
    \includegraphics[width=0.495\columnwidth]{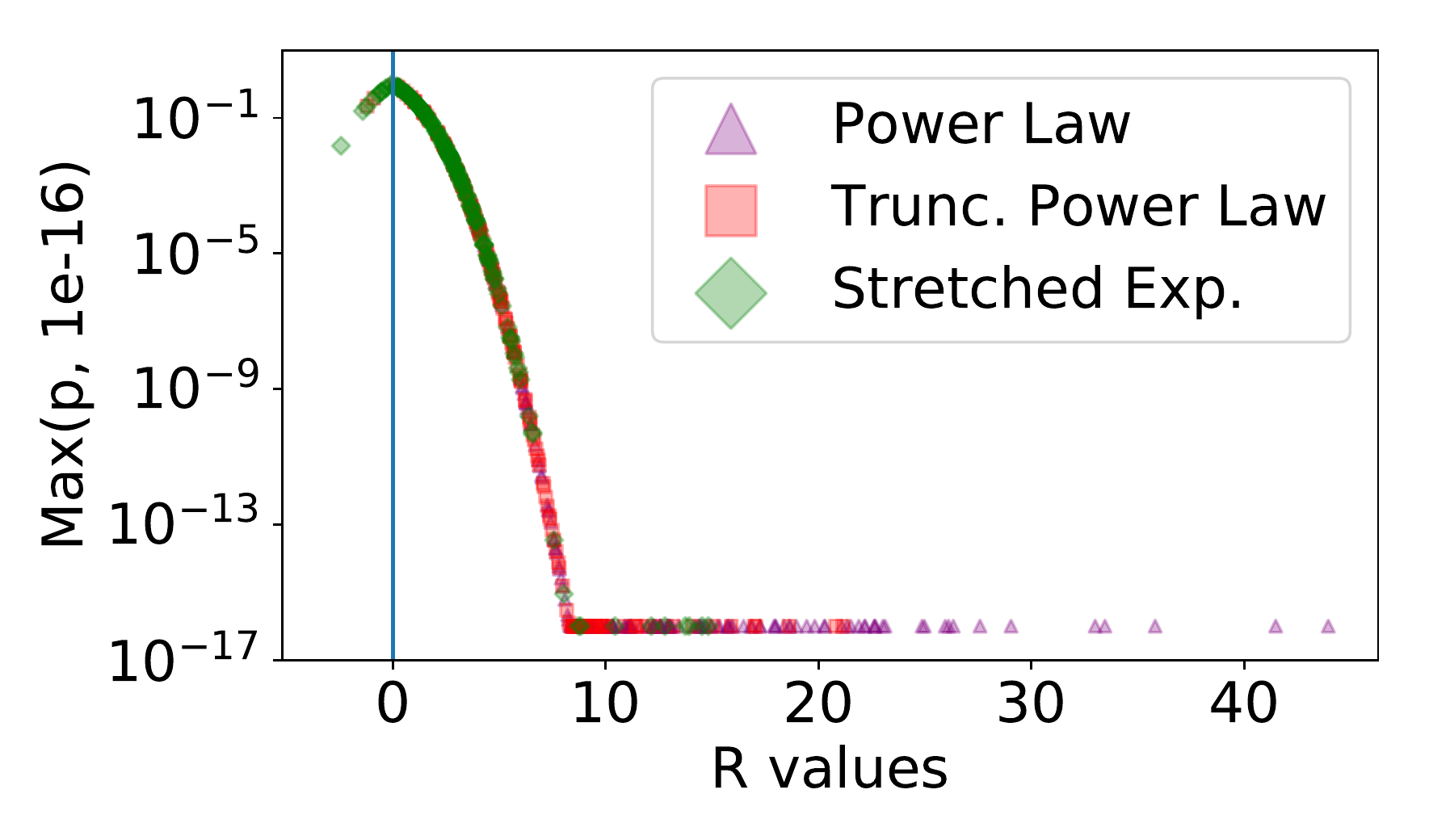}
    \caption{
    {\bf (Top Left)} The distribution of fitted lognormal parameters for tag frequency across 167 Stack Exchange communities
    (we exclude the outlier \textsc{patent}), which are approximately normally distributed.
    {\bf (Top Right)} The CDF of the KS Statistic (D) of fitted heavy-tailed distribution to the empirical data.
    The lognormal distribution has the smallest statistics, and $D < 0.06$ for 80\% of datasets;
    the only community with $D > 0.15$ is \textsc{patent}.
    {\bf (Bottom)} The $p$-values for comparing an alternative heavy tailed degree distribution hypothesis to the
    null hypothesis of the lognormal (left) and the distribution of the $p$ value with the log-likelihood ratio $R$ (right).
    The small $p$-values and positive log-likelihood ratios indicate that the lognormal is a better fit for
    the data compared to other common heavy-tailed distributions.
    }
    \label{fig:ks}
\end{figure}

More formally, we fit the parameters of a lognormal, power law,
truncated power law, and stretched exponential distributions to the tag frequency of each Stack Exchange community
using the \texttt{powerlaw} Python package~\cite{alstott2013powerlaw}. \Cref{fig:ks} (top left) shows the
fitted parameters, which are themselves approximately normally distributed.
We use two standard procedures for evaluating the fit of the lognormal: the Kolmogorov-Smirnov (KS) statistic
and the likelihood ratio test comparing the lognormal to other heavy-tailed degree distributions~\cite{power-law}.
The distribution of the KS statistics is much smaller for the lognormal compared to the other distributions (\cref{fig:ks}, top right)
and is less than 0.06 for 80\% of the Stack Exchange communities. Furthermore, the $p$-values from the likelihood ratio
test show that the power law, truncated power law, and stretched exponentials are not likely alternatives to the null
of a lognormal (\cref{fig:ks}, bottom).

To summarize, a lognormal distribution is an appropriate model for the distribution of tag frequencies.
In the next section, we describe a simple generative model for random bipartite graphs of tags and questions
based on this lognormal distribution. We will then later see that this model matches the real data in
a number of characteristics related to the co-tagging, i.e., how multiple tags are used on the same question.


\section{A Generative Model for Bipartite Tag-Question Networks}

In this section, we propose a simple generative model for the bipartite tag-question network.
Later, we will see that this model is able to recover many properties of the co-tagging network
of Stack Exchange communities, i.e, the graph where nodes correspond to tags, and edges connect
tags that have been applied to the same question. Formally, the bipartite tag-question graph $B$
consists of disjoint vertex sets $T$ and $Q$, each corresponding to the set of tags and questions,
as well as a set of undirected edges $E$; where $(t, q) \in E$ with $t \in T$ and $q \in Q$ signifies
that tag $t$ is applied to question $q$.
The frequency, or number of occurrences, of a tag $t$ is then simply the degree of $t$ in the graph $B$.

Our random network model has two basic steps. First, given $N_T=\lvert T \rvert$,
$N_Q = \lvert Q \rvert$, and the parameters $\mu$ and $\sigma$ of a
lognormal distribution, we first generate a sequence of tag occurrence counts
$x_t \sim \text{Lognormal}(\mu, \sigma^2)$.  These samples are scaled by a
constant so that $\sum_{t}x_t = m$ (where $m$ is the total number of tag
occurrences in the original dataset) and then rounded to an integer. Since scaling a
lognormal random variable by a constant is still lognormally distributed, we maintain
this property of the tag distribution, and this preserves the total number of tag-question pairs in the dataset.
Second, we assign tag $t$ to $x_t$ questions chosen uniformly at random without replacement. In this simplified
version of the model, the output deviates from the Stack Exchange networks in
two ways: (i) it is possible that a question has no tags and (ii) it is possible
that a question is assigned more than five tags. We now show how to account for
these deviations, and \cref{alg:generative} describes the full algorithm.

\begin{algorithm}[tb]
\SetAlgoLined
\DontPrintSemicolon
 \KwIn{number of tags $N_T$; number of questions $N_Q$;  \\
            target number of tag occurrences $m$; $\mu$, $\sigma^2$}
 \KwOut{tag-question bipartite graph $B = (T \cup Q, E)$}
 \tcc{Sample tag occurrences and compute corrections.}
 $x_t' \sim \text{LogNormal}(\mu,\sigma^2)$, $t = 1, \ldots, N_T$.\;
 $x_t \gets \text{round}(m \cdot x_t' / \sum_{t=1}^{N_T}x_t')$, $t = 1, \ldots, N_T.$\;
 Solve $\hat{N}_Q - \hat{N}_Q\exp(-m / \hat{N}_Q) = N_Q$ for $\hat{N}_Q$.\;
 $\hat{N}_Q$ $\gets$ round($\hat{N}_Q$).\;
  \tcc{Construct bipartite graph}
  $T \gets \{1, \ldots, N_T\}$, $Q \gets \{1, \ldots, \hat{N}_Q\}$.\;
  \For{each tag $t \in T$}{
    $Q_t \gets $ uniform sample of $x_t$ questions from $Q$.\;
    \lFor{$q \in Q_t$}{add edge $(t, q)$ to edge set $E$.}
}
$Q \gets \{q \in Q \;\vert\; \exists t \in T \text{ for which } (t, q) \in E\}$
\caption{Simple generative model for creating random bipartite graphs of tags and questions.}
\label{alg:generative}
\end{algorithm}

\subsubsection{Correction for question counts}
To fix the problem where questions can have
no tags, we make a ``correction'' in the number of questions. More specifically,
we increase the number of questions from $N_Q$ to $\hat{N}_Q$ so that after the
random assignment, the expected number of questions with at least one tag is
close in expectation to $N_Q$, the number of questions in the empirical dataset.
We then simply discard questions with no tags (\cref{alg:generative}).

We approximate the expected number of questions with no tags under a
simplification where tags can be duplicated in questions (the approximation is
not necessary, but it makes the calculations simpler, has small variance
theoretically, and provides good results empirically).  Here, the probability
that a question gets $0$ tags is the same for each question---it is just the
probability that all tags are assigned to the other $\hat{N}_Q - 1$ questions:
\begin{align*}
\prod_{i=1}^{N_T}\prod_{j=0}^{x_i-1} \left[1- 1/(\hat{N}_Q-j)\right] \approx (1- 1/\hat{N}_Q)^m,
\end{align*}
where $m$ is total number of tag occurrences.
Thus, since $\hat{N}_Q$ and $m$ are generally large, when assigning tags uniformly at random to $\hat{N}_Q$ questions,
the expected number of questions with 0 tags is
\[
\hat{N}_Q(1- 1 / \hat{N}_Q)^{m} \approx \hat{N}_Q\exp(-m / \hat{N}_Q).
\]
There are $N_Q$ questions if the following equation is satisfied:
\begin{equation}\label{eqn:correction}
\hat{N}_Q - \hat{N}_Q\exp(-m / \hat{N}_Q) = N_Q.
\end{equation}
We claim that \cref{eqn:correction} has a unique positive solution $\hat{N}_Q > N_Q$.
Since $m$ and $N_Q$ are positive constants, the left hand side of \cref{eqn:correction}
is a function $f$ of $\hat{N}_Q$.
Moreover, the function $f$ is
continuous and monotonically increasing in $\hat{N}_Q$, and $f(N_Q) =  N_Q(1-\exp(-m/N_Q)) < N_Q$.
Therefore, the above equation has a unique positive
solution for $\hat{N}_Q$ that is larger than $N_Q$. We can find the solution
efficiently with binary search, and then round $\hat{N}_Q$ to the
nearest integer.

\begin{figure}
    \centering
    \includegraphics[width=0.495\columnwidth]{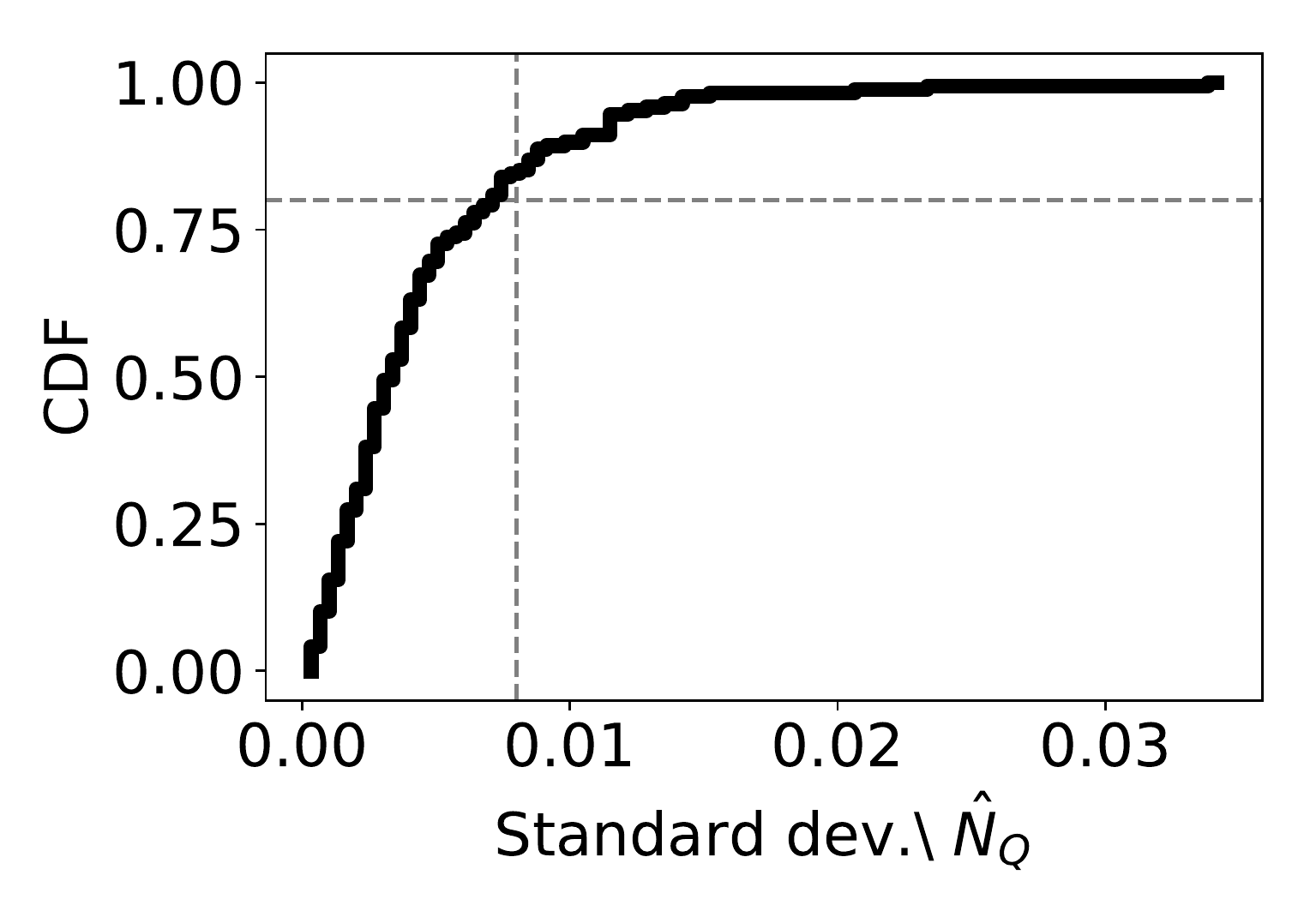}
     \includegraphics[width=0.495\columnwidth]{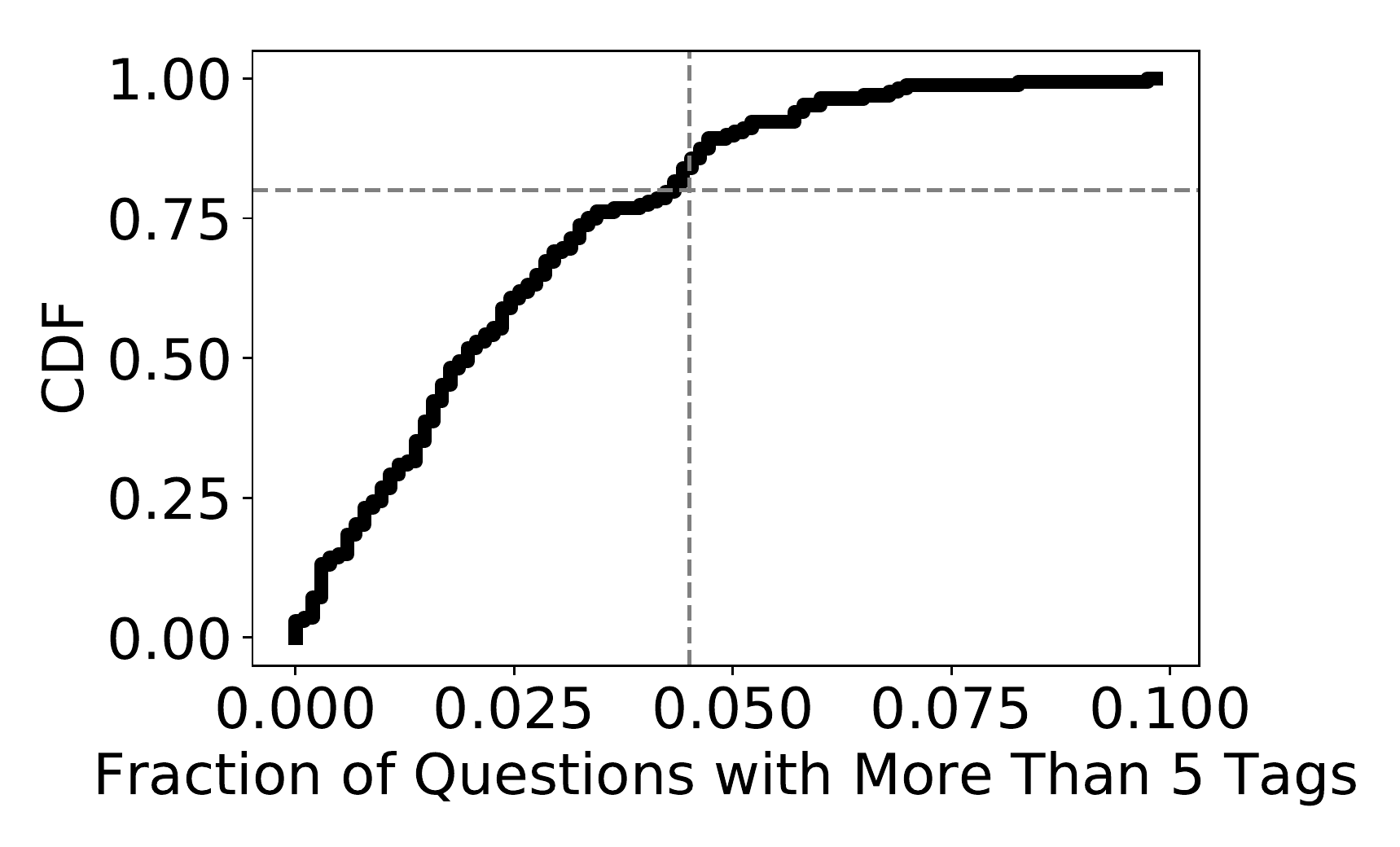}
    \caption{{\bf (Left)} The CDF of the ratio of the (theoretical) standard deviations to the corrected number of questions.
    The small ratio (less than 0.008 for 80\% of datasets) shows that our correction for the number of questions is accurate.
    {\bf (Right)} The CDF of the fraction of questions with more than five tags in one sample of the random graph for each dataset.
    This fraction is small---less than 0.045 for 80\% of datasets, which justifies the relaxation in our random graph model.
    }
    \label{fig:var}
\end{figure}

In our experiments, using the corrected number of questions with our model is
accurate, even with our approximations.  Generating one sample for each dataset,
the relative error between the number of questions with at least one tag in the
model deviates from the true number of questions by 0.32\% on average and by at
most 3.75\% across all datasets.  While these statistics are for just one sample
in each network, the variance in the number of questions with 0 tags is
approximately $\hat{N}_Qp(1-p)$. The ratio between the theoretical standard deviation
and the corrected number of questions is small----less than 0.008 for 80\% of the datasets
(\cref{fig:var}, left).

\begin{figure}
\centering
\includegraphics[width=0.495\columnwidth]{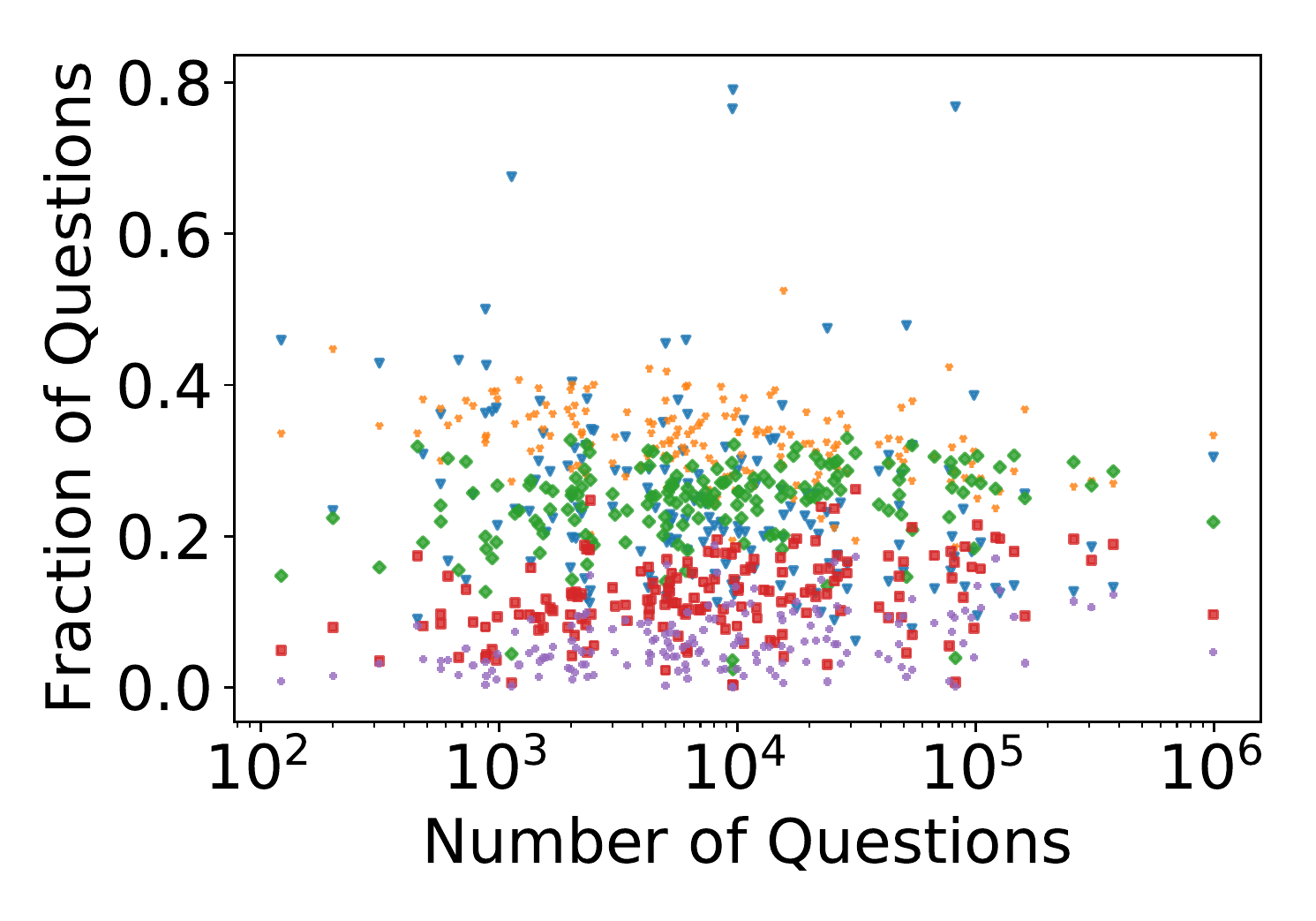}
\includegraphics[width=0.495\columnwidth]{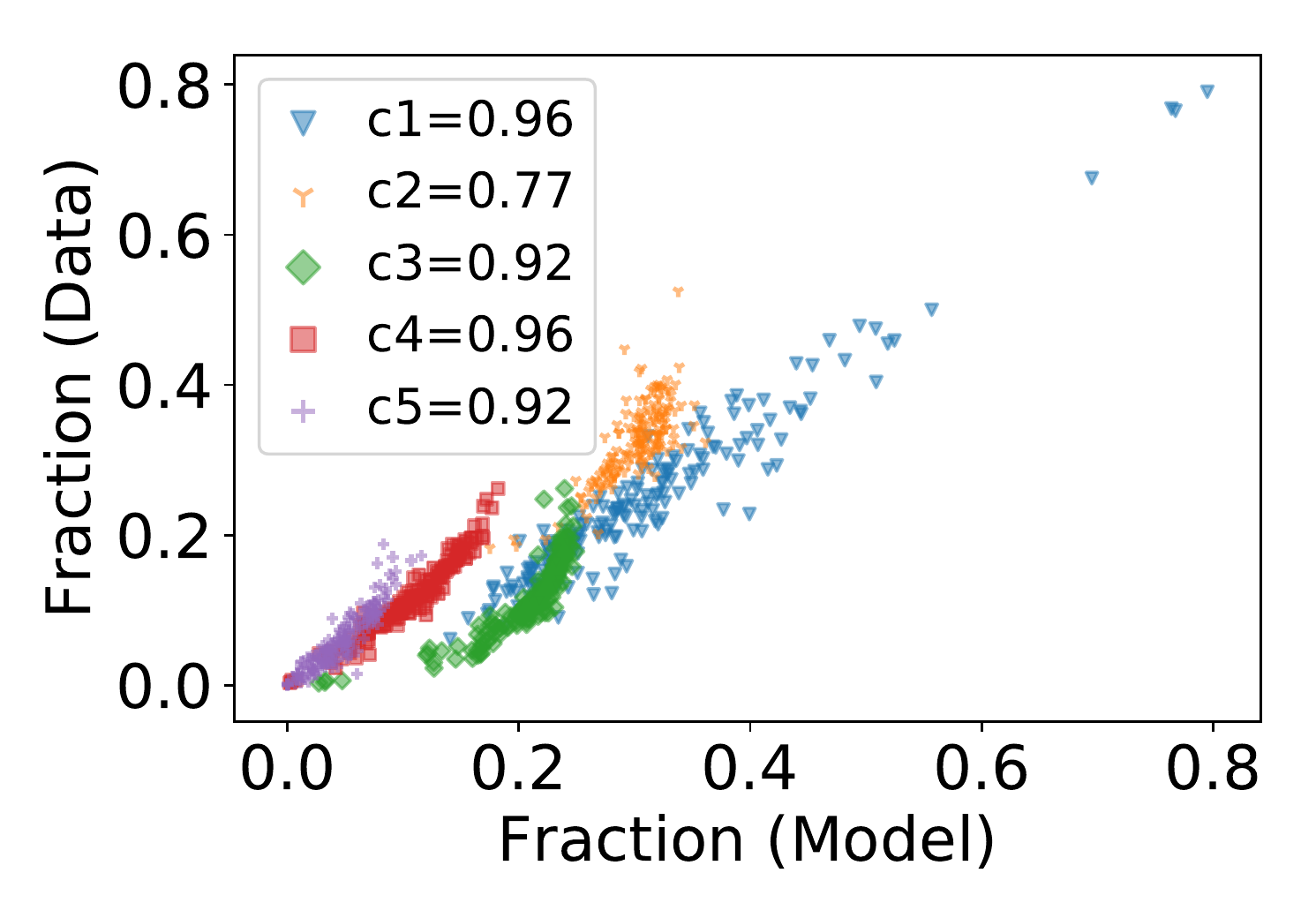}
\caption{Distributions of number of tags per post. Markers indicate number of tags: 1---blue triangle; 2---yellow `Y'; 3---green diamond;
4---red square; and 5---purple `+'.
{\bf (Left)} Fraction of questions with a given number of tags as a function of the number of questions in a datasets.
The distribution of tags per post is roughly independent of the size of the Stack Exchange community.
{\bf (Right)} Comparison of the distribution of the number of tags per post in the data and a sample from our generative model.
The distributions are strongly correlated (shown in legend).
}
\label{fig:num_tag_pquestion}
\end{figure}

\subsubsection{Number of tags per question}
We next justify our second model deviation, which is that questions can be
assigned more than five tags.  Our argument is that only a small fraction of
questions are actually assigned more than five tags with our generative model.
We generated tag-question bipartite graphs with \cref{alg:generative} for each
Stack Exchange community using the fitted lognormal parameters (\cref{fig:ks},
bottom).  The mean fraction of questions with more than 5 tags in the generated
networks across 168 Stack Exchange platforms is only 2.5\% and more than 80\% of
datasets have less than a 4.5\% of questions with more than five tags
(\cref{fig:var}, right).

\subsubsection{Summary}
\Cref{alg:generative} is a simple generative model for bipartite tag-question
networks that generates tag occurrences with the lognormal distribution that we
found to be common across nearly all Stack Exchange communities. As a first look at how our
model matches the empirical data, we consider the distribution of the number of
tags per question. In the empirical data, this distribution tends to be uncorrelated
with the size of the dataset (\cref{fig:num_tag_pquestion}, left). We also find that
the distribution of the number of tags per question in the model closely matches the
empirical data (\cref{fig:num_tag_pquestion}, right). In the next section, we
analyze co-tagging, i.e., how tags jointly annotate questions.  Our model has no
built-in notion of correlations in co-tagging, yet we find that the model still
matches macroscopic co-tagging properties in the data.


\section{Co-tagging Analysis}
In addition to the bipartite tag-question network, we also build a ``co-tagging network''
for each Stack Exchange community.
Recall that the tag-question network $B = (T \cup Q, E)$ is given by vertex sets $T$ and $Q$
corresponding to tags and questions and has edges $(t, q) \in E$ connecting tags
to questions. The co-tagging network $G$ is the projection of this graph onto the
set of tags. Formally, $G = (T, F)$, where $(s, t) \in F$ if and only if there is some
question $q \in Q$ such that $(s, q), (t, q) \in E$. In this case, we say that $s$ and $t$
\emph{co-tag} with each other. We also associate a weight with each edge in $G$ corresponding
to the number of questions containing the two tags (the number of times that two nodes are co-tagged):
\begin{equation}\label{eq:weights}
w_{s, t} = \lvert \{q \in Q \;\vert\; (s, q), (t, q) \in E\} \rvert.
\end{equation}

In the rest of this section, we show that co-tagging networks constructed from samples of our generative model (\cref{alg:generative})
match statistical properties of the co-tagging networks of empirical data,
even though our model does not explicitly account for co-tagging behavior.
Again, we use the lognormal parameters $\mu$ and $\sigma$ fitted for each dataset (\cref{fig:ks}) to generate a random graph
for each Stack Exchange network.
We focus our attention on three properties of the co-tagging network:
(i) the expected number of co-tags (i.e., the weighted degree in $G$) as a function of tag frequency;
(ii) the expected number of unique co-tags (i.e., the unweighted degree in $G$), again as a function of tag frequency; and
(iii) weighted and unweighted versions of the clustering coefficient of the graph $G$.

\begin{figure}
\centering
\includegraphics[width=0.495\columnwidth]{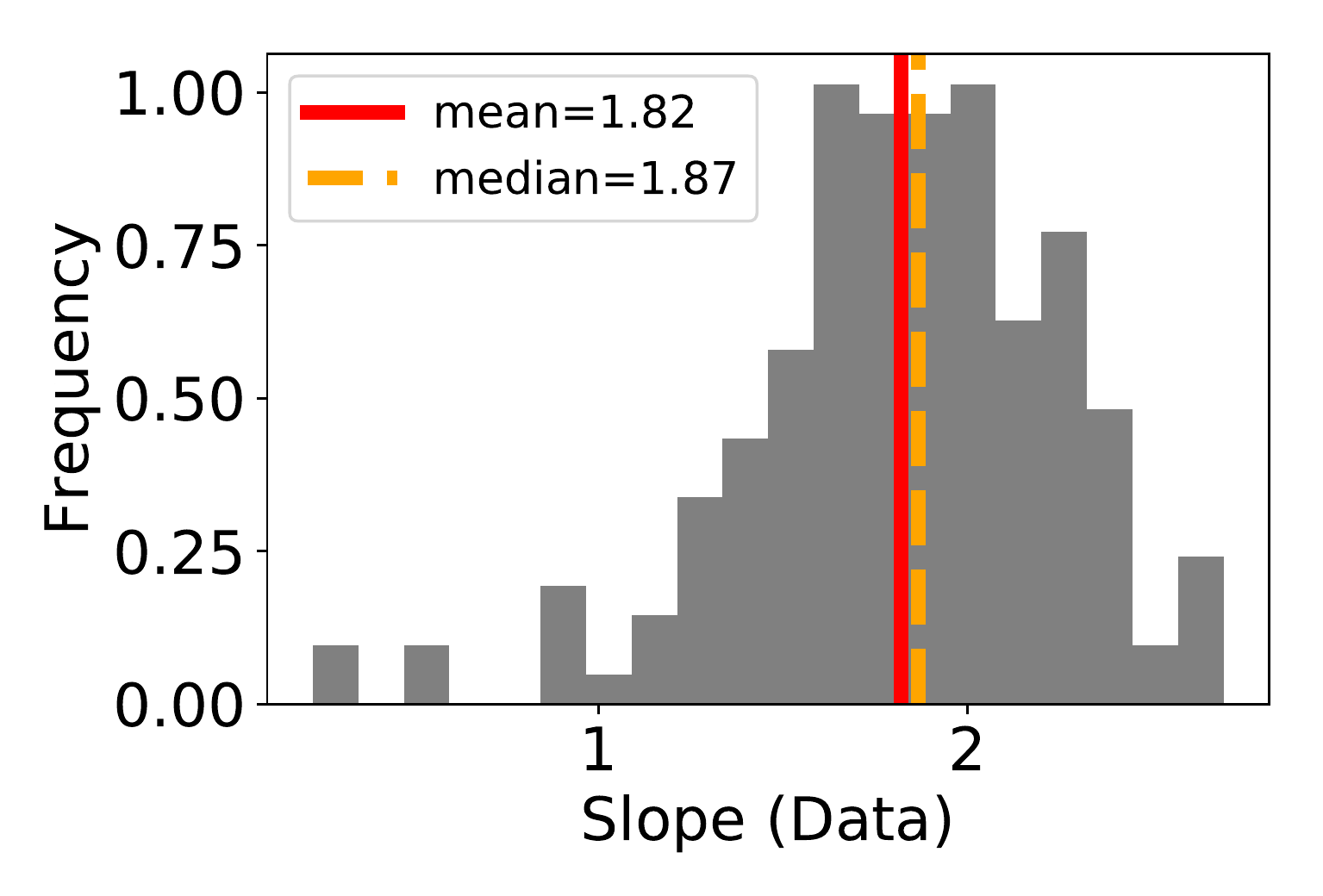}
\includegraphics[width = 0.495\columnwidth]{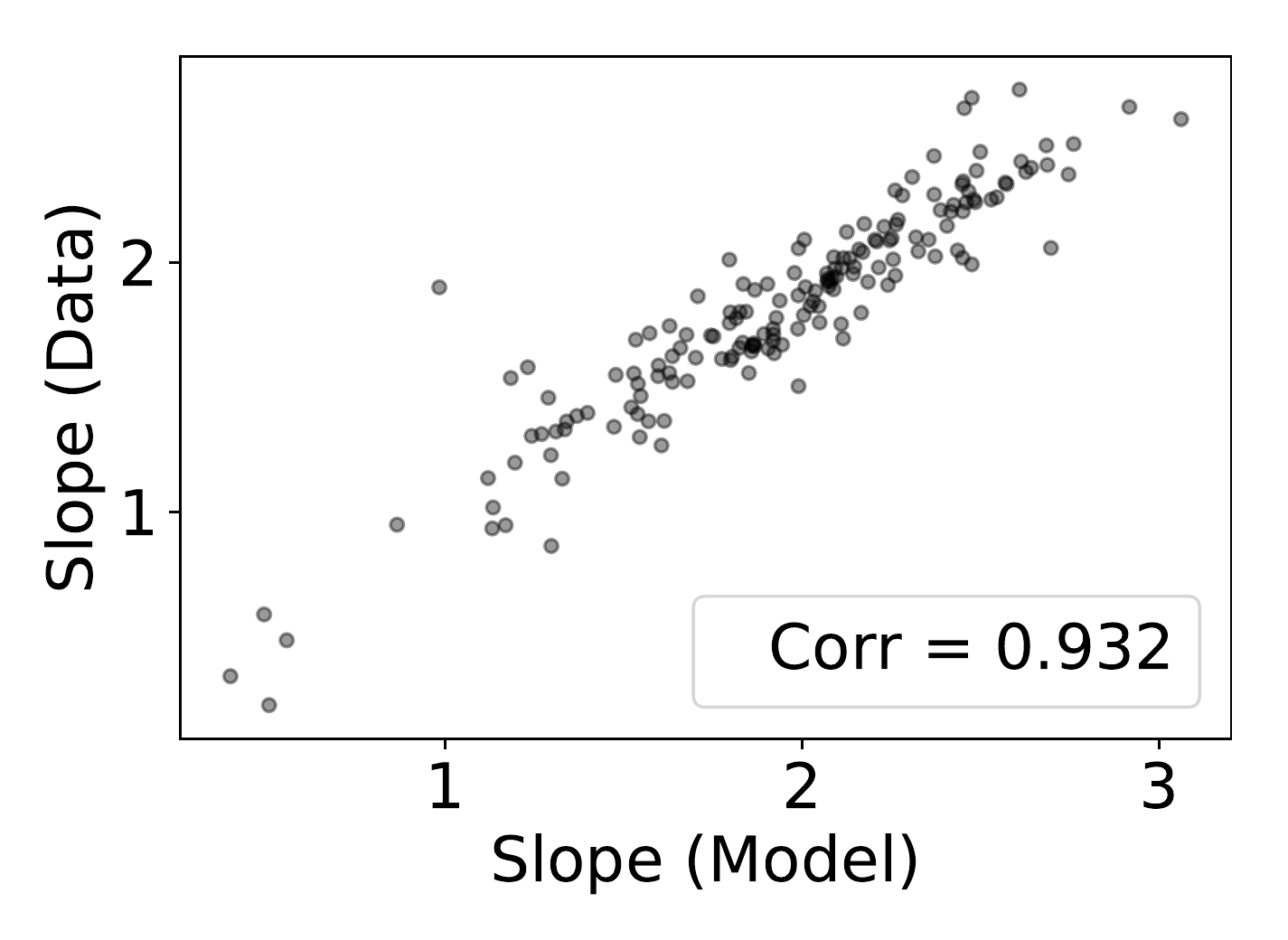}
\caption{{\bf (Left)} The weighted number of co-tags is approximately a linear function of tag frequency.
Here, we show the distribution of slopes from the linear regression
over our collection of Stack Exchange communities. The regression has an $r^2$ value greater than 0.95 in 95\%
of the empirical datasets and greater than 0.97 in 97\% of the generated datasets.
{\bf (Right)} The relationship between the fitted slope on the data and in the model across the Stack Exchange communities,
which are strongly correlated.}
\label{fig:linear}
\end{figure}

\subsection{Weighted Co-tags and Tag Frequency}
We first examine the relationship between the number
of co-tags of a given tag as a function of its frequency (the number of questions in which it appears).
Here, we consider the number of co-tags to be weighted, i.e.,
the number of co-tags of tag $t$ is $k_t = \sum_{s \in T} w_{s, t}$, following \cref{eq:weights}.
In the empirical data, this relationship is essentially linear---a
linear model of the number of co-tags in regressed on the number of questions containing the tag
has a coefficient of determination ($r^2$ value) greater than 0.95 in 95\% of the Stack Exchange communities.
\Cref{fig:linear} (left) shows the distribution of the slopes, which concentrate around 1.82.

We now show why we would also expect this behavior from our model.
Recall that the generative model samples tag frequencies $x_t \sim \text{LogNormal}(\mu,\sigma^2)$ and
then scaled so that these variables to match the total number of tag occurrences.
The number of co-tags between $s$ and $t$ then follows $w_{s,t} \sim \text{Hypergeom}(\hat{N}_Q, x_s, x_t)$,
where $\hat{N}_Q$ is the corrected number of questions in \cref{alg:generative}.
Thus, the expected number of co-tags $k_t$ of a given tag $t$ is
\[
\mathbb{E}[k_t]  = \sum_{t\neq s, s=1,...,N_T} \frac{x_s x_t}{\hat{N}_Q} = \frac{(m - x_t)x_t}{\hat{N}_Q},
\]
where $m$ is the target number of tag occurrences (the first equality
comes from the independence in assignment of the tags).
Although there is a quadratic relationship between $k_t$ and $x_t$,
we know that $x_t$ is typically small compared to $m$. Thus, the
gradient is well-approximated by the linear function $m / \hat{N}_Q$,
i.e., $\frac{d}{dx_t}k_t \approx m / \hat{N}_Q$, independent of $x_t$.
Our analysis here is independent of the lognormal distribution of the tag
frequency---we only relied on independence in the way
that tags are assigned to questions.

In actual random samples, the linear relationship holds.
We performed the same linear regression on random samples
from our generative model using the fitted parameters in \cref{fig:ks}
as we did for the empirical datasets.
In the model, 97\% of the 168 datasets have a correlation coefficient $r^2 > 0.97$.
Furthermore, the slopes from the regression on the generated data are highly
correlated with the slopes on the empirical data (the correlation is 0.932; \cref{fig:linear}, right),
and the mean squared error between the slope derived from a sample from the generative model
and the computed slope on the empirical data across all Stack Exchange communities is just 0.10.

\subsection{Unique Co-tags and Tag Frequency}
In the above analysis, we saw that the number of co-tags of a given tag is approximately linear in the number
of questions in which the tag appears---in both the empirical data and our model-generated data.
In this section, we instead consider the number of \emph{unique}
co-tags of a given tag $t$ as a function of the number of questions containing tag $t$.
In this case, the number of unique co-tags is equal to the unweighted degree of tag $t$
in the co-tagging network $G$ defined above.

\begin{figure}
    \centering
    \includegraphics[width = 0.49\columnwidth]{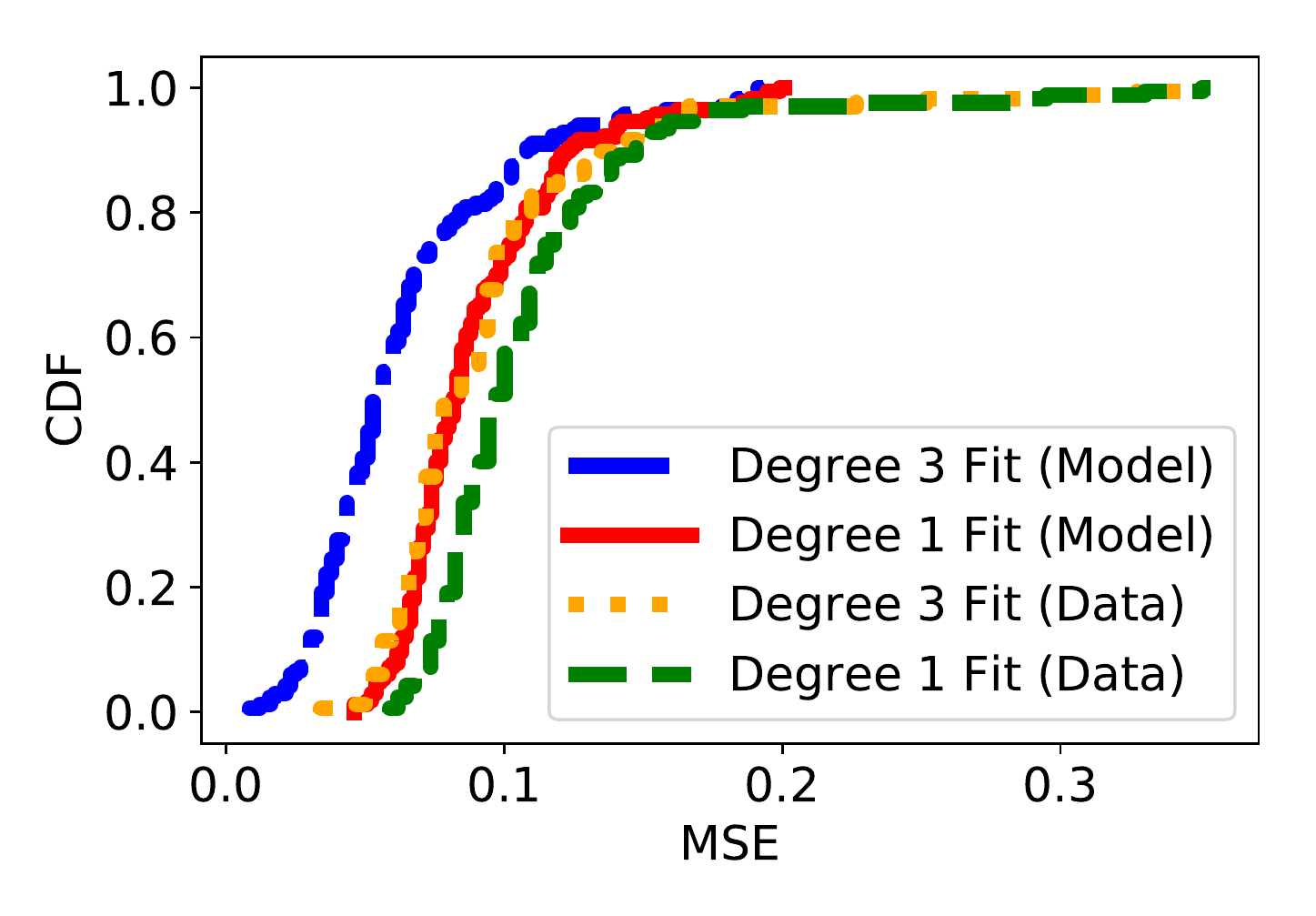}
    \includegraphics[width = 0.49\columnwidth]{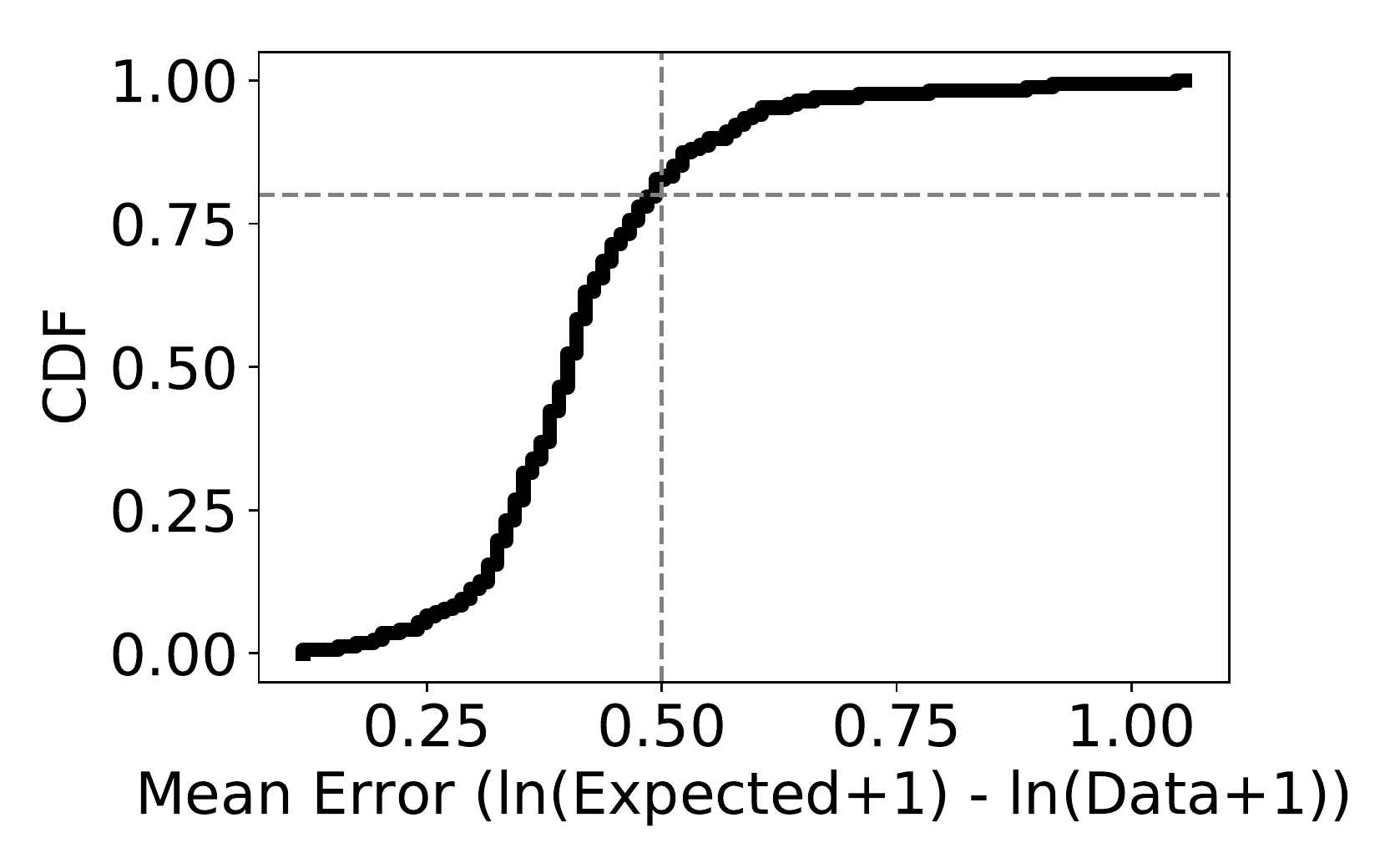}
    \caption{{\bf (Left)} The CDF of the mean-squared error in third-degree and first-degree (linear) polynomial models of
    of the log number of unique co-tags in terms of the log of tag frequency in both the data and the samples from the generative model.
    The third-degree polynomial is a good approximation and matches the expected value of the model (\cref{fig:question_uniquec}).
    {\bf{(Right)}} The CDF of the mean error in the expected number of unique co-tags in the model and the
    actual number of co-tags in the dataset. The error is less than 0.5 in 80\% of the datasets. The model slightly over-estimates the number
    of unique co-tags by not taking into account tag correlations (see also \cref{fig:question_uniquec}).
     }
    \label{fig:mse}
\end{figure}

 \begin{figure}
    \centering
    \includegraphics[width=\columnwidth]{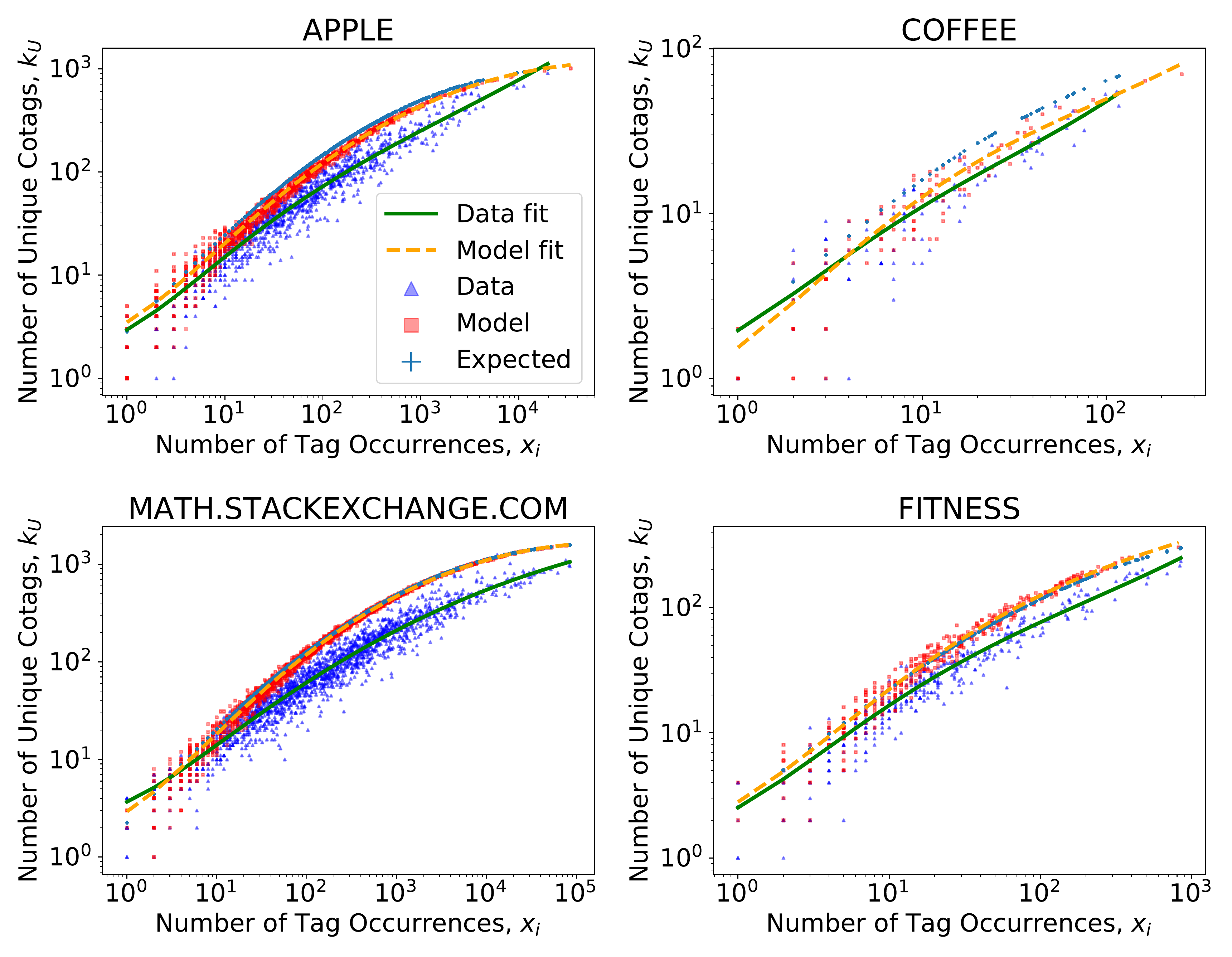}
    \caption{Relationship between the number of unique co-tags and tag frequency
    on four Stack Exchange communities, which is well-approximated
    by a degree-three polynomial (see also \cref{fig:mse}).
    The model has the same shape, albeit slightly above the data.}
    \label{fig:question_uniquec}
\end{figure}

We find that the log of the number of unique co-tags is well-approximated as a third-degree polynomial
of the log of the number of question that contain the tag. Formally, let $d_t$ denote the unweighted
degree of tag $t$ in the co-tagging network $G$ and $x_t$ the number of questions containing tag $t$.
We then fit a the following polynomial model:
\begin{equation}\label{eqn:pol}
\log(d(t) + 1) = \sum_{i=0}^3 a_i \log(x_t + 1)^i.
\end{equation}
\Cref{fig:mse} (left) shows the CDF of the mean-squared error of the polynomial fit.
The third-degree polynomial is a good fit for both the empirical data and the model
across the collection of Stack Exchange communities.
\Cref{fig:question_uniquec} shows the distributions and fit of the third-degree polynomial for a
few representative networks.
In these cases, the polynomial fit is accurate and captures the fact that the number
of unique co-tags does not grow linearly with tag frequency.
Instead, the growth in unique co-tags tapers for some of the most frequently used tags.
This happens because there is a limited total number of tags (\cref{fig:data_sta}), so tags that occur
frequently have fewer options to increase the number of unique co-tags.

Interestingly, the fitted third-degree polynomial coefficients $\{a_i\}$, when taken as a collection across the Stack Exchange communities,
largely lie on a lower-dimensional subspace. In the empirical datasets,
the first principal component explains 86\% of the variability,
and the second principal component explains an additional 13\% of the variability.
Similar results hold for the fitted coefficients in datasets generated with our
model---89\% of the variability is explained with the first principal component and an addition 10\%
is explained by the second principal component.

We can easily compute the expected number of unique co-tags
with a simple summation.
We argued in the previous section that the weighted number of
co-tags between tags $s$ and $t$ is $w_{s,t} \sim \text{Hypergeom}(\hat{N}_Q, x_s, x_t)$. Thus,
the expected number of unique co-tags $d_t$ of tag $t$ is
\begin{align*}
\mathbb{E}[d_t]
= \sum_{s \neq t} \mathbb{P}(w_{s, t} > 0)
&= \sum_{s \neq t} 1 - \mathbb{P}(w_{s, t}  = 0) \\
&= \sum_{s \neq t} \left[1 - \frac{\binom{\hat{N}_Q - x_s}{x_t}}{\binom{\hat{N}_Q}{x_t}}\right],
\end{align*}
where $x_s$ is the sampled number of questions for tag $s$ in \cref{alg:generative}
and $\hat{N}_Q$ is the corrected number of questions.
\Cref{fig:question_uniquec} shows that the generated model data matches this expectation.



\begin{figure}
\centering
\includegraphics[width=1.0\columnwidth]{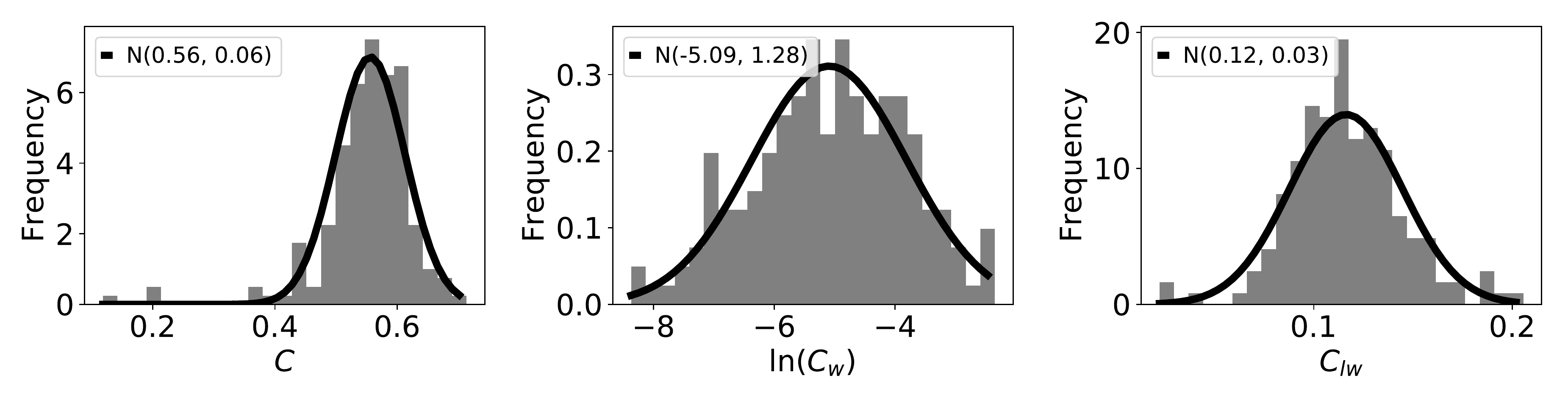}
\includegraphics[width=1.0\columnwidth]{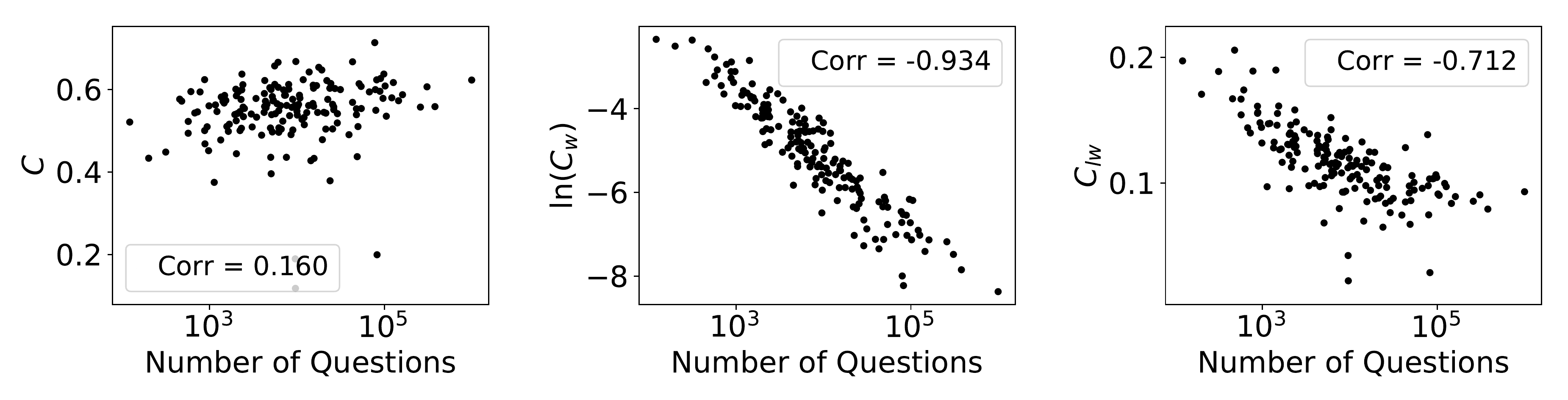}
\includegraphics[width = 1.0\columnwidth]{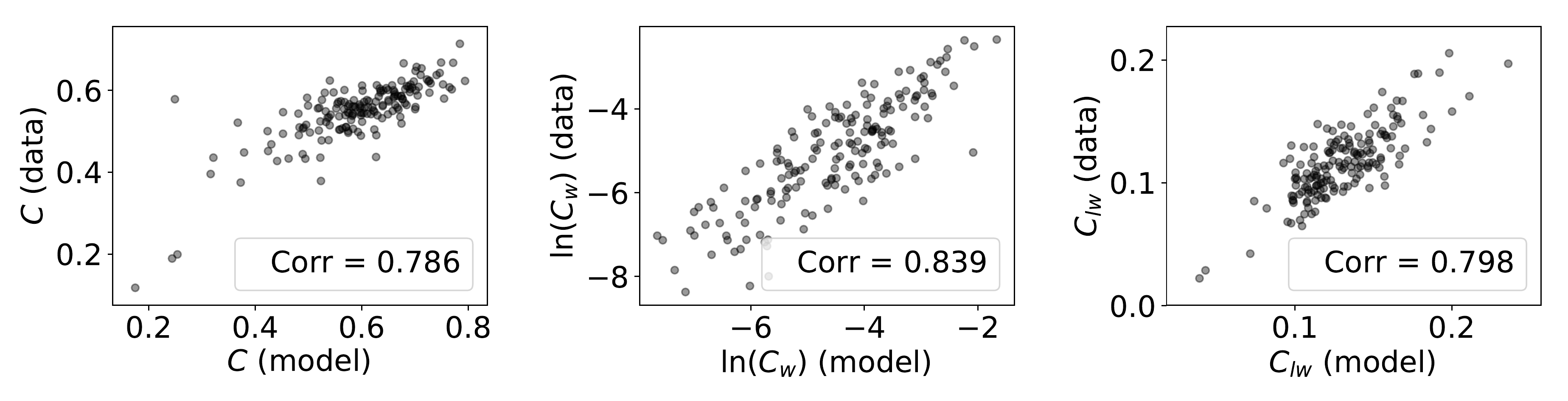}
\caption{{\bf (Top)} The clustering coefficients
of the Stack Exchange communities are approximately normally distributed.
{\bf (Middle)} The unweighted clustering coefficient $C$ has a weak correlation with respect to the size
of the community as measured by the log-number of questions; the weighted versions are negative correlated.
{\bf (Bottom)} The clustering coefficients in the co-tagging networks generated by our model
are similar to the clustering coefficients of the empirical Stack Exchange communities.
}
\label{fig:cc_data}
\end{figure}


\subsection{Clustering in the co-tagging networks}
Finally, we analyze the clustering coefficient of the co-tagging networks,
which is one of the fundamental measurements in networks~\cite{Watts1998Collective,Newman-2003-survey}.
Let $\Delta_u$, $d_u$, and $w_{u,v}$ be
the number of triangles containing node $u$,
the unweighted degree of $d_u$,
and the weight of edge $(u, v)$ in the co-tagging graph $G$.
We consider three clustering coefficients:
\begin{enumerate}
\item The unweighted clustering coefficient~\cite{Watts1998Collective}:
\[
C = \frac{1}{\lvert T \rvert}\sum_{u \in T}\frac{2 \Delta_u}{d_u(d_u-1)}
\]
\item The weighted clustering clustering coefficient:
\[
C_w = \frac{1}{\lvert T \rvert}\sum_{u \in T}\frac{1}{d_u(d_u-1)} \sum_{v,z} (\hat{w}_{u,v} \hat{w}_{u,z} \hat{w}_{v,z})^{1/3},
\]
where $\hat{w}_{u,v} = w_{u,v}/\max_{x,y}w_{x,y}$~\cite{weighted_cc}.
We will analyze $\log(C_w)$.
\item The log-weighted clustering coefficient, which is the same as the mean weighted clustering coefficient,
except the weight $w_{u,v}$ is replaced by $w'_{u,v} = \log(w_{u,v} + 1)$:
\[
C_{lw} = \frac{1}{\lvert T \rvert}\sum_{u \in T}^n\frac{1}{d_u(d_u-1)} \sum_{v,z} (\hat{w}'_{u,v} \hat{w}'_{u,z} \hat{w}'_{v,z})^{1/3},
\]
where $\hat{w}'_{u,v} = w'_{u,v}/\max_{x,y}w'_{x,y}$ and summations over cases where $w'_{u,v} = 0$ (i.e., with no edge)
are ignored.
\end{enumerate}

\Cref{fig:cc_data} (top row) shows that all three clustering coefficients are approximately normally distributed across
the collection of Stack Exchange communities.
Furthermore, the unweighted coefficients are only weakly correlated with the size of the community,
measured by the log-number of questions on the Stack Exchange (\cref{fig:cc_data}, middle row).
We conclude that the size of a Stack Exchange community is likely not a driving factor in the unweighted clustering of the network,
which backs up conventional wisdom for the analysis of real-world networks~\cite{Newman-2003-survey},
differs from the behavior of random graph models that produce heavy-tailed degree distributions, where clustering decreases with size~\cite{Bollobas-2004-results}.
On the other hand, the weighted clustering coefficients tend to decrease with the size of the Stack Exchange community
(\cref{fig:cc_data}, middle row).

The co-tagging networks derived from samples of our generative model reproduce
these clustering coefficients remarkably closely and with strong positive
correlations (\cref{fig:cc_data}, bottom row).  Again, we emphasize that our
model does not bake in any explicit notion of clustering. Instead, our model
only matches the lognormal distribution of the tag frequency and the total
number of tags applied to all questions. Thus, clustering in the co-tagging
in Stack Exchange communities could be explained simply by these
simpler statistics. This finding contrasts sharply with typical (social) network analysis, where
clustering is exhibited at a much higher level than is expected by random graph
models~\cite{Newman-2003-survey}. The key difference is that our model is based on a projection
of a bipartite tag-question graph rather than directly modeling the co-tagging network.
This type of modeling has a long history in sociology~\cite{Breiger-1974-duality} but has
received relatively less theoretical attention in social network analysis~\cite{Lattanzi-2009-affiliation}.


\section{Discussion}

In addition to providing answers to questions, the users of Q\&A platforms
create knowledge through annotation of questions. With their tagging system,
Stack Exchange provides a unique opportunity to study these annotations for two
main reasons. First, tags cannot be created freely and there are community
guidelines for their application, which differs substantially from tagging norms on other social media platforms.
Second, there is a collection of Stack Exchange communities that have largely evolved independently, enabling us to
model and analyze tagging with more statistical evidence. And we indeed found
similarities in macroscopic tagging structure---in terms of tag frequency and
co-tagging network structure---across 168 Stack Exchange communities
spanning a diverse range of topics. This contrasts from typical network analyses
that study a single snapshot of a social network. Previously, researchers have
circumvented this issue by looking at, for example,
sets of disparate subgraphs from a larger graph~\cite{Traud-2012-FB100,Patania-2017-collaborations};
samples of ego networks~\cite{Ugander-2013-subgraphs,Benson-2018-Closure,Mcauley-2014-ego}; and
collections of snapshots of time-evolving networks~\cite{Yaverolu-2014-revealing}.

One macroscopic property across communities is that the distribution of tag
frequencies is well-modeled by a lognormal distribution.
The fitted parameters of the lognormal distributions (\cref{fig:ks}) themselves are
approximately normally distributed across our collection of Stack Exchange communities.
Thus, one could incorporate this information as a simple prior in bayesian modeling of tag-question networks.

We used the tag frequency distribution to develop a simple generative model for random tag-question bipartite graphs,
which was able to reproduce a number of the co-tagging and clustering properties of the
datasets, without explicitly modeling correlations or clustering in the co-tagging
process. Further understanding of the process producing this distribution is an
avenue for future research. For example, multiplicative growth models are a
well-known generative process for lognormal
distributions~\cite{Mitzenmacher-2004-history}. Although outside the scope of
this paper, the availability of temporal information from Stack Exchange
provides a path towards more robust understanding of the underlying processes of
tag use, similar to other methods for estimating growth on the Web and in social
networks~\cite{Huberman-1999-WWW,Overgoor-2018-choosing}.

\xhdr{Code and data}
Code to reproduce our results, along with processed data,
are available at \url{https://github.com/yushangdi/stack-exchange-cotagging}.

\xhdr{Acknowledgments}
This research was supported by NSF award DMS-1830274
and ARO award 86798.


\bibliography{main}

\begin{thebibliography}{}

\bibitem[\protect\citeauthoryear{Adamic \bgroup et al\mbox.\egroup
  }{2008}]{yahoo1}
Adamic, L.~A.; Zhang, J.; Bakshy, E.; and Ackerman, M.~S.
\newblock 2008.
\newblock Knowledge sharing and yahoo answers: Everyone knows something.
\newblock In {\em Proceedings of the 17th International Conference on World
  Wide Web (WWW)},  665--674.

\bibitem[\protect\citeauthoryear{Akoglu, Chandy, and
  Faloutsos}{2013}]{Akoglu-2013-opinion}
Akoglu, L.; Chandy, R.; and Faloutsos, C.
\newblock 2013.
\newblock Opinion fraud detection in online reviews by network effects.
\newblock In {\em International AAAI Conference on Web and Social Media
  (ICWSM)}, volume~13,  2--11.

\bibitem[\protect\citeauthoryear{Alstott, Bullmore, and
  Plenz}{2013}]{alstott2013powerlaw}
Alstott, J.; Bullmore, E.; and Plenz, D.
\newblock 2013.
\newblock Powerlaw: a python package for analysis of heavy-tailed
  distributions.

\bibitem[\protect\citeauthoryear{Anderson \bgroup et al\mbox.\egroup
  }{2012}]{Anderson-2012-value}
Anderson, A.; Huttenlocher, D.; Kleinberg, J.; and Leskovec, J.
\newblock 2012.
\newblock Discovering value from community activity on focused question
  answering sites: a case study of {Stack Overflow}.
\newblock In {\em Proceedings of the 18th {ACM} {SIGKDD} International
  Conference on Knowledge Discovery and Data Mining (KDD)}.

\bibitem[\protect\citeauthoryear{Anderson \bgroup et al\mbox.\egroup
  }{2013}]{Anderson-2013-badges}
Anderson, A.; Huttenlocher, D.; Kleinberg, J.; and Leskovec, J.
\newblock 2013.
\newblock Steering user behavior with badges.
\newblock In {\em Proceedings of the 22nd International Conference on World
  Wide Web (WWW)}.

\bibitem[\protect\citeauthoryear{Bascompte, Jordano, and
  Olesen}{2006}]{Bascompte-2006-coevolution}
Bascompte, J.; Jordano, P.; and Olesen, J.~M.
\newblock 2006.
\newblock Asymmetric coevolutionary networks facilitate biodiversity
  maintenance.
\newblock {\em Science} 312(5772):431--433.

\bibitem[\protect\citeauthoryear{Benson \bgroup et al\mbox.\egroup
  }{2018}]{Benson-2018-Closure}
Benson, A.~R.; Abebe, R.; Schaub, M.~T.; Jadbabaie, A.; and Kleinberg, J.
\newblock 2018.
\newblock Simplicial closure and higher-order link prediction.
\newblock {\em Proceedings of the National Academy of Sciences}
  115(48):E11221--E11230.

\bibitem[\protect\citeauthoryear{Bollob{\'{a}}s and
  Riordan}{2004}]{Bollobas-2004-results}
Bollob{\'{a}}s, B., and Riordan, O.~M.
\newblock 2004.
\newblock Mathematical results on scale-free random graphs.
\newblock In {\em Handbook of Graphs and Networks}. Wiley.
\newblock  1--34.

\bibitem[\protect\citeauthoryear{Bosu \bgroup et al\mbox.\egroup }{2013}]{rep1}
Bosu, A.; Corley, C.~S.; Heaton, D.; Chatterji, D.; Carver, J.~C.; and Kraft,
  N.~A.
\newblock 2013.
\newblock Building reputation in {StackOverflow}: An empirical investigation.
\newblock In {\em Working Conference on Mining Software Repositories},  89--92.

\bibitem[\protect\citeauthoryear{Breiger}{1974}]{Breiger-1974-duality}
Breiger, R.~L.
\newblock 1974.
\newblock The duality of persons and groups.
\newblock {\em Social Forces} 53(2).

\bibitem[\protect\citeauthoryear{Capocci and Caldarelli}{2008}]{citeulike}
Capocci, A., and Caldarelli, G.
\newblock 2008.
\newblock Folksonomies and clustering in the collaborative system citeulike.
\newblock {\em Journal of Physics A: Mathematical and Theoretical} 41(22).

\bibitem[\protect\citeauthoryear{Cattuto \bgroup et al\mbox.\egroup
  }{2007}]{Cattuto-2007-folksonomy}
Cattuto, C.; Schmitz, C.; Baldassarri, A.; Servedio, V. D.~P.; Loreto, V.;
  Hotho, A.; Grahl, M.; and Stumme, G.
\newblock 2007.
\newblock Network properties of folksonomies.
\newblock {\em AI Communications} 20(4):245--262.

\bibitem[\protect\citeauthoryear{Cattuto \bgroup et al\mbox.\egroup
  }{2009}]{Cattuto-2009-folksonomy}
Cattuto, C.; Barrat, A.; Baldassarri, A.; Schehr, G.; and Loreto, V.
\newblock 2009.
\newblock Collective dynamics of social annotation.
\newblock {\em Proceedings of the National Academy of Sciences}
  106(26):10511--10515.

\bibitem[\protect\citeauthoryear{Chen, Ho, and Kim}{2010}]{Chen-2010-Google}
Chen, Y.; Ho, T.-H.; and Kim, Y.-M.
\newblock 2010.
\newblock Knowledge market design: A field experiment at google answers.
\newblock {\em Journal of Public Economic Theory} 12(4):641--664.

\bibitem[\protect\citeauthoryear{Chojnacki and
  K{\l}opotek}{2010}]{Chojnacki-2010-folksonomy}
Chojnacki, S., and K{\l}opotek, M.
\newblock 2010.
\newblock Random graph generative model for folksonomy network structure
  approximation.
\newblock {\em Procedia Computer Science} 1(1):1683--1688.

\bibitem[\protect\citeauthoryear{Clauset, Shalizi, and
  Newman}{2009}]{power-law}
Clauset, A.; Shalizi, C.; and Newman, M.
\newblock 2009.
\newblock Power-law distributions in empirical data.
\newblock {\em SIAM Review} 51(4):661--703.

\bibitem[\protect\citeauthoryear{Goh \bgroup et al\mbox.\egroup
  }{2007}]{Goh-2007-disease}
Goh, K.-I.; Cusick, M.~E.; Valle, D.; Childs, B.; Vidal, M.; and Barabasi,
  A.-L.
\newblock 2007.
\newblock The human disease network.
\newblock {\em Proceedings of the National Academy of Sciences}
  104(21):8685--8690.

\bibitem[\protect\citeauthoryear{Halpin, Robu, and Shepherd}{2007}]{cotag1}
Halpin, H.; Robu, V.; and Shepherd, H.
\newblock 2007.
\newblock The complex dynamics of collaborative tagging.
\newblock In {\em Proceedings of the 16th International Conference on World
  Wide Web (WWW)}.

\bibitem[\protect\citeauthoryear{Huberman and Adamic}{1999}]{Huberman-1999-WWW}
Huberman, B.~A., and Adamic, L.~A.
\newblock 1999.
\newblock Growth dynamics of the world-wide web.
\newblock {\em Nature} 401(6749):131--131.

\bibitem[\protect\citeauthoryear{Kwak \bgroup et al\mbox.\egroup
  }{2010}]{Kwak-2010-Twitter}
Kwak, H.; Lee, C.; Park, H.; and Moon, S.
\newblock 2010.
\newblock What is {Twitter}, a social network or a news media?
\newblock In {\em Proceedings of the 19th International Conference on World
  Wide Web (WWW)}.

\bibitem[\protect\citeauthoryear{Larremore, Clauset, and
  Jacobs}{2014}]{Larremore-2014-biSBM}
Larremore, D.~B.; Clauset, A.; and Jacobs, A.~Z.
\newblock 2014.
\newblock Efficiently inferring community structure in bipartite networks.
\newblock {\em Physical Review E} 90(1).

\bibitem[\protect\citeauthoryear{Lattanzi and
  Sivakumar}{2009}]{Lattanzi-2009-affiliation}
Lattanzi, S., and Sivakumar, D.
\newblock 2009.
\newblock Affiliation networks.
\newblock In {\em Proceedings of the 41st annual {ACM} Symposium on Theory of
  Computing (STOC)}.

\bibitem[\protect\citeauthoryear{MacLeod}{2014}]{exper1}
MacLeod, L.
\newblock 2014.
\newblock Reputation on {Stack Exchange}: Tag, you're it!
\newblock In {\em 28th International Conference on Advanced Information
  Networking and Applications Workshops},  670--674.

\bibitem[\protect\citeauthoryear{Maity, Sahni, and Mukherjee}{2015}]{topic_pop}
Maity, S.; Sahni, J. S.~S.; and Mukherjee, A.
\newblock 2015.
\newblock {Analysis and Prediction of Question Topic Popularity in Community
  Q\&A Sites: A Case Study of Quora}.
\newblock In {\em International AAAI Conference on Web and Social Media
  (ICWSM)}.

\bibitem[\protect\citeauthoryear{Mcauley and Leskovec}{2014}]{Mcauley-2014-ego}
Mcauley, J., and Leskovec, J.
\newblock 2014.
\newblock Discovering social circles in ego networks.
\newblock {\em {ACM} Transactions on Knowledge Discovery from Data} 8(1):1--28.

\bibitem[\protect\citeauthoryear{Mitzenmacher}{2004}]{Mitzenmacher-2004-history}
Mitzenmacher, M.
\newblock 2004.
\newblock A brief history of generative models for power law and lognormal
  distributions.
\newblock {\em Internet Mathematics} 1(2):226--251.

\bibitem[\protect\citeauthoryear{Newman}{2003}]{Newman-2003-survey}
Newman, M. E.~J.
\newblock 2003.
\newblock The structure and function of complex networks.
\newblock {\em {SIAM} Review} 45(2):167--256.

\bibitem[\protect\citeauthoryear{Onnela \bgroup et al\mbox.\egroup
  }{2005}]{weighted_cc}
Onnela, J.~P.; Saramäki, J.; Kertész, J.; and Kaski, K.
\newblock 2005.
\newblock Intensity and coherence of motifs in weighted complex networks.
\newblock {\em Physical Review E}.

\bibitem[\protect\citeauthoryear{Overgoor, Benson, and
  Ugander}{2018}]{Overgoor-2018-choosing}
Overgoor, J.; Benson, A.~R.; and Ugander, J.
\newblock 2018.
\newblock Choosing to grow a graph: Modeling network formation as discrete
  choice.
\newblock {\em arXiv:1811.05008}.

\bibitem[\protect\citeauthoryear{Pal, Chang, and
  Konstan}{2012}]{Pal-2012-evolution}
Pal, A.; Chang, S.; and Konstan, J.~A.
\newblock 2012.
\newblock Evolution of experts in question answering communities.
\newblock In {\em International AAAI Conference on Web and Social Media
  (ICWSM)}.

\bibitem[\protect\citeauthoryear{Paranjape, Benson, and
  Leskovec}{2017}]{Paranjape-2017-motifs}
Paranjape, A.; Benson, A.~R.; and Leskovec, J.
\newblock 2017.
\newblock Motifs in temporal networks.
\newblock In {\em Proceedings of the Tenth {ACM} International Conference on
  Web Search and Data Mining}.

\bibitem[\protect\citeauthoryear{Patania, Petri, and
  Vaccarino}{2017}]{Patania-2017-collaborations}
Patania, A.; Petri, G.; and Vaccarino, F.
\newblock 2017.
\newblock The shape of collaborations.
\newblock {\em {EPJ} Data Science} 6(1).

\bibitem[\protect\citeauthoryear{Paul, Hong, and Chi}{2012}]{rep2}
Paul, S.~A.; Hong, L.; and Chi, E.~H.
\newblock 2012.
\newblock Who is authoritative? understanding reputation mechanisms in {Quora}.
\newblock In {\em Collective Intelligence}.

\bibitem[\protect\citeauthoryear{Posnett \bgroup et al\mbox.\egroup
  }{2012}]{stackE}
Posnett, D.; Warburg, E.; Devanbu, P.; and Filkov, V.
\newblock 2012.
\newblock Mining {Stack Exchange}: Expertise is evident from initial
  contributions.
\newblock In {\em International Conference on Social Informatics},  199--204.

\bibitem[\protect\citeauthoryear{Salganik}{2006}]{Salganik-2006-MusicLab}
Salganik, M.~J.
\newblock 2006.
\newblock Experimental study of inequality and unpredictability in an
  artificial cultural market.
\newblock {\em Science} 311(5762):854--856.

\bibitem[\protect\citeauthoryear{Tian, Zhang, and Li}{2013}]{tian2013towards}
Tian, Q.; Zhang, P.; and Li, B.
\newblock 2013.
\newblock Towards predicting the best answers in community-based
  question-answering services.
\newblock In {\em International AAAI Conference on Web and Social Media
  (ICWSM)}.

\bibitem[\protect\citeauthoryear{Traud, Mucha, and
  Porter}{2012}]{Traud-2012-FB100}
Traud, A.~L.; Mucha, P.~J.; and Porter, M.~A.
\newblock 2012.
\newblock Social structure of {Facebook} networks.
\newblock {\em Physica A: Statistical Mechanics and its Applications}
  391(16):4165--4180.

\bibitem[\protect\citeauthoryear{Ugander, Backstrom, and
  Kleinberg}{2013}]{Ugander-2013-subgraphs}
Ugander, J.; Backstrom, L.; and Kleinberg, J.
\newblock 2013.
\newblock Subgraph frequencies: mapping the empirical and extremal geography of
  large graph collections.
\newblock In {\em Proceedings of the 22nd International Conference on World
  Wide Web (WWW)}.

\bibitem[\protect\citeauthoryear{Ugander \bgroup et al\mbox.\egroup
  }{2011}]{Ugander-2011-anatomy}
Ugander, J.; Karrer, B.; Backstrom, L.; and Marlow, C.
\newblock 2011.
\newblock The anatomy of the facebook social graph.
\newblock {\em arXiv:1111.4503}.

\bibitem[\protect\citeauthoryear{Vander~Wal}{2005}]{vander}
Vander~Wal, T.
\newblock 2005.
\newblock Folksonomy.
\newblock \url{http://www.vanderwal.net/random/entrysel.php?blog=1622}.

\bibitem[\protect\citeauthoryear{Wang \bgroup et al\mbox.\egroup
  }{2013}]{Wang2013}
Wang, G.; Gill, K.; Mohanlal, M.; Zheng, H.; and Zhao, B.~Y.
\newblock 2013.
\newblock Wisdom in the social crowd: An analysis of quora.
\newblock In {\em Proceedings of the 22nd International Conference on World
  Wide Web (WWW)},  1341--1352.

\bibitem[\protect\citeauthoryear{Wang, Liu, and Fan}{2011}]{cotag2}
Wang, X.; Liu, H.; and Fan, W.
\newblock 2011.
\newblock Connecting users with similar interests via tag network inference.
\newblock In {\em Proceedings of the 20th ACM International Conference on
  Information and Knowledge Management},  1019--1024.

\bibitem[\protect\citeauthoryear{Watts and
  Strogatz}{1998}]{Watts1998Collective}
Watts, D.~J., and Strogatz, S.~H.
\newblock 1998.
\newblock Collective dynamics of `small-world' networks.
\newblock {\em Nature} 393(6684):440--442.

\bibitem[\protect\citeauthoryear{Yang \bgroup et al\mbox.\egroup
  }{2011}]{yang2011culture}
Yang, J.; Morris, M.~R.; Teevan, J.; Adamic, L.~A.; and Ackerman, M.~S.
\newblock 2011.
\newblock Culture matters: A survey study of social {Q\&A} behavior.
\newblock In {\em International AAAI Conference on Web and Social Media
  (ICWSM)}.

\bibitem[\protect\citeauthoryear{Yavero{\u{g}}lu \bgroup et al\mbox.\egroup
  }{2014}]{Yaverolu-2014-revealing}
Yavero{\u{g}}lu, O.~N.; Malod-Dognin, N.; Davis, D.; Levnajic, Z.; Janjic, V.;
  Karapandza, R.; Stojmirovic, A.; and Pr{\v{z}}ulj, N.
\newblock 2014.
\newblock Revealing the hidden language of complex networks.
\newblock {\em Scientific Reports} 4(1).

\end{thebibliography}
\bibliographystyle{aaai}
\end{document}